\documentclass[draft,onecolumn]{IEEEtran}
\usepackage{stmaryrd}
\usepackage{amssymb}
\usepackage{mathrsfs}
\usepackage{amsfonts}  
\usepackage[dvips,final]{graphicx}
\usepackage[dvips]{color}
\usepackage{epsfig,amssymb,amsmath}
\newcommand{\beq}{\begin{equation}}
\newcommand{\eeq}{\end{equation}}
\newcommand{\beqn}{\begin{eqnarray}}
\newcommand{\eeqn}{\end{eqnarray}}

\begin{document}

\listoffigures

\newpage

\begin{center}

{\Large\bf MIMO Radar Using Compressive Sampling \footnote{Copyright
\copyright 2008 IEEE. Personal use of this material is permitted. However,
permission to use this material for any other purposes must be
obtained from the IEEE by sending a request to
pubs-permissions@ieee.org. \\

This work
was supported by the by the Office of Naval Research under Grant
ONR-N-00014-07-1-0500, and the National Science Foundation under
Grants  CNS-06-25637 and CNS-04-35052 }}

\bigskip
{\it Yao Yu and Athina P. Petropulu} \\
          Department of Electrical \& Computer Engineering,
          Drexel University, Philadelphia, PA 19104\\
          \medskip
          {\it H. Vincent Poor}\\
          School of Engineering and Applied Science,
          Princeton University, Princeton, NJ 08544
\end{center}

\bigskip

\begin{abstract}
A  MIMO radar system is proposed for obtaining angle and Doppler
information on potential targets. Transmitters  and receivers are
nodes of a small scale wireless network and are assumed to be
randomly scattered on a disk. The transmit nodes transmit
uncorrelated waveforms.
Each receive node applies compressive sampling to the received signal to obtain a small number of samples, which the node subsequently forwards
to a fusion center.
Assuming that the targets are sparsely located in the
angle-Doppler space, based on the samples forwarded by the receive nodes the fusion center formulates an
$\ell_1$-optimization problem,  the solution of which yields target
angle and Doppler information.
 The proposed approach achieves the
superior resolution of MIMO radar with far fewer samples than
required by other approaches. This implies power savings during the
communication phase between the receive nodes and  the fusion center.
Performance in the presence of a jammer is analyzed for the case of
slowly moving targets. Issues related to forming the basis matrix
that spans the angle-Doppler space, and for selecting a grid for
that space are discussed. Extensive simulation results are provided to demonstrate the performance of the proposed approach at difference jammer and noise levels.


\end{abstract}
\section{Introduction}

 Multiple-input multiple-output
(MIMO) radar systems have received considerable recent attention,
e.g., \cite{Fishler:04}-\cite{Li:07}. Unlike a conventional transmit
beamforming radar system that uses highly correlated waveforms, a
MIMO radar system transmits multiple independent waveforms via its
antennas. A MIMO radar system is advantageous in two different
scenarios \cite{Haimovich: 08}-\cite{Chen:081}. In the first one
\cite{Haimovich: 08}, the transmit antennas are located far apart
from each other relative to their distance to the target. {This
enables the radar to view the target from different directions
simultaneously. The MIMO radar system transmits independent probing
signals from decorrelated transmitters through different paths, and
thus each target return carries independent information about the
target. Combining these independent target returns results in  a
diversity gain, which
 enables the MIMO radar system to reduce  target
radar cross section (RCS) scintillations and achieve high target
resolution.  In the second scenario \cite{Stoica: 07m}, a MIMO radar
is equipped with $M_t$ transmit and $N_r$ receive antennas that are
close to each other relative to the target, so that the RCS does not
vary between the different paths. In this scenario, the phase
differences induced by
 transmit and receive
antennas can  be exploited to form a long virtual array with
$M_tN_r$ elements. This enables the MIMO radar system to achieve
superior spatial resolution as compared  to a traditional radar
system.
 MIMO radar can achieve  a desired beampattern by transmitting correlated  waveforms
\cite{Stoica:07}-\cite{Fuhrmann:08}. This is useful in cases where
the radar system wishes to avoid certain directions because they
either correspond to eavesdroppers, or are known to be of no
interest. In this paper we consider closely spaced transmit and
receive antennas and uncorrelated transmit waveforms.

Compressive sampling (CS) \cite{Donoho:06}-\cite{Candes:08} has
received considerable attention recently, and has been applied
successfully in diverse fields, e.g., image processing
\cite{Romberg:08} and wireless communications
\cite{Bajwa:06}\cite{Paredes:07}. The theory of CS states that a
$K$-sparse signal $\mathbf{x}$ of length $N$ can be recovered
exactly with high probability from $\mathcal{O}(K\log N)$
measurements via $\ell_1$-optimization. Let $\Psi$ denote the basis
matrix that spans this sparse space, and let $\Phi$ denote a
measurement matrix. The convex optimization problem arising from CS
is formulated as follows
\begin{eqnarray}
\min\|\mathbf{s}\|_1,\ \  s.t. \ \text{to}\ {\bf y}=\Phi{\bf x}
=\Phi \Psi {\bf s}
\end{eqnarray}
where $\mathbf{s}$ is a sparse vector with $K$ principal elements
and the remaining elements can be ignored;  $\Phi$ is an $M\times N$
matrix with $M\ll N$, that is incoherent with  $\Psi$.

The application of  compressive sampling to a radar system was
recently investigated in \cite{Baraniuk:07}, \cite{Gurbuz:07} and
\cite{Herman:08}. In \cite{Baraniuk:07}, in the context of radar
imaging, compressive sampling was shown to have the potential to
reduce the typically required sampling rate and even render matched
filtering unnecessary.
 In \cite{Gurbuz:07}, a  CS-based data
acquisition and imaging algorithm for ground penetrating radar was
proposed to exploit the sparsity of targets in the spatial
dimension. The approach of \cite{Gurbuz:07} was shown to require
fewer measurements  than standard backprojection methods. In
\cite{Herman:08}, CS was applied in a radar system with a  small
number of targets, exploiting target sparseness in the
time-frequency shift plane. The work of \cite{Gurbuz:08} considered
direction of arrival (DOA) estimation of signal sources using CS.
Although \cite{Gurbuz:08} focussed  on  communication systems, the
proposed approach can be straightforwardly extended to radar
systems. { In \cite{Gurbuz:08}, the basis matrix $\Psi$ was formed
by the discretization of the angle space. The source signals were
assumed to be unknown, and an approximate version of the basis
matrix was obtained based on the signal received by a reference
vector. The signal at the reference sensor would have to be sampled
at a very high rate  in order to construct a good basis matrix.

In this paper, we consider a small scale network that acts as a MIMO
radar system. Each node is equipped with one antenna, and the nodes
are  distributed at random on a disk of a certain radius. Without
any fixed infrastructure, the distributed antennas in this small
network render such MIMO radar more flexible  than a fixed antenna
array since we can choose the nodes freely.  For example, the
network nodes could be soldiers that carry antennas on their
backpacks. We refer to such a  MIMO radar system a distributed MIMO
radar. The nodes transmit independent waveforms. We extend the idea
of \cite{Gurbuz:08} to the problem of angle-Doppler estimation for
MIMO radar.  Since the number of targets is typically smaller than
the number of snapshots that can be obtained, angle-Doppler
estimation can be formulated as that of recovery of a sparse vector
using CS. Unlike the scenario considered in \cite{Gurbuz:08}, in
MIMO radar the transmitted waveforms are known at each  receive
node. This information, and also information on the location of
transmit nodes, if available, enables each receive node to construct
the basis matrix locally, without knowledge of the received signal
at a reference sensor or any other antenna.
In cases in which the
receive nodes do not have location information about the
transmitters, or they do not have the computational
power, or they face significant interference,  the received samples are transmitted to a fusion center which
has access to location information and also to computational power. Based on the received data, the fusion center
formulates an augmented $\ell_1$-optimization problem the solution of which provides target angle and Doppler information. The performance
of $\ell_1$-optimization  depends on the noise level. A potential
jammer would act as noise, and thus affect performance. We
 provide analytical expressions for the average
signal-to-jammer ratio (SJR) and propose a modified measurement
matrix that improves the SJR. For the case of stationary targets,
the proposed approach is compared to existing methods, such as the
Capon, amplitude and phase estimation (APES), generalized likelihood
ratio test (GLRT) \cite{Xu:06} and multiple signal classification
(MUSIC) methods, while for moving targets, comparison to
  the matched filter method \cite{Levanon:04} is conducted.

Preliminary results of our work were published in
\cite{Petropulu:08}. Independently derived results  for MIMO radar
using compressive sampling were also published in the same
proceedings \cite{Chen:08}. The difference between our work and
\cite{Chen:08} is that in \cite{Chen:08} a uniform linear array was
considered as a transmit and receive antenna configuration, while in
our work we focus on randomly placed transmit and receive antennas,
i.e., an infrastructure-less  MIMO radar system. Further we study
the effects of a jammer on estimation performance.

 The paper is organized as follows. In Section II we provide the
signal model of a distributed MIMO radar system.  In Section III,
the proposed approach for angle-Doppler estimation is presented. In
Section IV we derive  the average SJR for the proposed approach and
also discuss a modification of the  random measurement matrix that
can further improve the SJR. Simulation results  are given in
Section V for the cases of stationary targets and moving targets.
Finally, we make some concluding remarks in Section VI.

\emph{Notation}: Lower case and capital letters in bold denote
respectively vectors and matrices.  The  expectation of a random
variable is denoted by $E\{\cdot\}$. The superscript
 $ (\cdot)^{H}$ and $\mathrm{Tr}(\cdot)$ denote respectively the
Hermitian transpose and  trace  of a matrix.

\section{Signal Model for MIMO Radar}\label{sig_model}
 We consider a MIMO
radar system with $M_t$ transmit nodes and $N_r$ receive nodes that
are uniformly distributed on a disk of  a small radius $r$.
This particular assumption  will be used  in Section
\ref{analysis on SJR} for the analytical evaluation of the proposed approach.
 For
simplicity, we assume that targets and nodes lie on the same plane
and we consider a clutter-free environment. Perfect synchronization
and localization of nodes is also assumed.
 The extension to the case
in which targets and nodes lie in 3-dimension space is
straightforward. Let $(r^t_i, \alpha^t_i)$ and $(r^r_i, \alpha^r_i)$
denote the locations in polar coordinates of the $i$-th transmit and
receive antenna, respectively. Then the probability density
functions of $r^{t/r}_i$ and $\alpha^{t/r}_i$ are}
\begin{eqnarray}
f_{r_i^{t/r}}(r_i^{t/r})&=&\frac{2r_i^{t/r}}{r^2}, \
0<r_i^{t/r}<r\nonumber \\
\mathrm{and} \ f_{\alpha_i^{t/r}}(\alpha_i^{t/r})&=&\frac{1}{2\pi},\
-\pi\leq\alpha_i^{t/r}<\pi. \label{2}
\end{eqnarray}

Let us assume that there are $K$ point targets present. The $k$-th
target is at azimuth angle $\theta_k$ and moves  with constant
radial speed $v_k$. Its range equals $d_k(t)=d_k(0)-v_kt$, where
$d_k(0)$ is the distance between this target and  the origin at time
equal to zero. Under the far-field assumption, i.e.,
 $d_{k}(t) \gg r^{t/r}_i$, the distance between the $i$th transmit/receive
 antenna and the $k$-th target
 $d^t_{ik}$/$d^r_{ik}$ can be approximated as
\begin{eqnarray}
d^{t/r}_{ik}(t) \approx d_k (t)-
\eta_i^{t/r}(\theta_k)=d_k(0)-v_kt-\eta_i^{t/r}(\theta_k)
\end{eqnarray}
where
$\eta_i^{t/r}(\theta_k)=r^{t/r}_i\cos(\theta_k-\alpha^{t/r}_i)$.

 Let $x_i(t)e^{j2\pi ft}$ denote the
continuous-time waveform transmitted by the $i$-th transmit antenna,
where $f$ is the carrier frequency; we assume that all transmit
nodes use the same carrier frequency and also that  the $x_i(t)$ is
periodic with period $T$ and narrowband. Besides, we also assume the
slowly moving targets, i.e., $\frac{v_k}{c}\ll 1$.

 The received signal at the $k$-th target equals
\begin{eqnarray}
y_{k}(t)&=& \beta_k\sum_{i=1}^{M_t} x_i(t-d^{t}_{ik}(t)/c)
\exp({j{2\pi}f(t-\frac{d^t_{ik}(t)}{c}) }), \ k=1,\ldots, K
\end{eqnarray}
where $\{\beta_k,k=1,\ldots,K\}$ are complex amplitudes proportional
to the RCS and are assumed to be the same for all the receivers. The
latter assumption is consistent with
 a small network in which  the distances between  network nodes are
much smaller than the distances between the  nodes and the targets,
i.e., $d_{k}(t) \gg r^{t/r}_i$. Thus, since they are closely spaced,
all receive nodes
 see the same aspect of the target.

Due to reflection by the target, the $l$-th antenna element receives
\begin{eqnarray}
z_{l}(t)&=&\sum_{k=1}^{K}y_k(t-\frac{d^r_{lk}(t)}{c})+\epsilon_{l}(t)\nonumber\\
&=&\sum_{k=1}^{K}\beta_k\sum_{i=1}^{M_t}
x_i(t-\frac{d^t_{ik}(t)+d^r_{lk}(t)}{c})e^{j{2\pi}f(t-\frac{d^t_{ik}(t)+d^r_{lk}(t)}{c})
}+\epsilon_{l}(t),\ l=1,\ldots, M_r
\end{eqnarray}
where $\epsilon_{l}(t)$ represents noise, which is assumed to be
independent and identically distributed  (i.i.d.) Gaussian with zero
mean and variance  $\sigma^2$.

The narrowband assumption on the transmit waveforms allows us to
ignore the delay in $x_i(t)$, and consider the delay in the phase
term  only.   Thus,  the received baseband signal at the $l$-th
antenna can be approximated as
\begin{eqnarray}
z_{l}(t)& \approx &\sum_{k=1}^{K}\beta_k\sum_{i=1}^{M_t}
x_i(t)e^{j{2\pi}f_kt}e^{j\frac{2\pi}{\lambda}(-2d^k(0)+\eta_i^{t}(\theta_k)+\eta_l^{r}(\theta_k))}+\epsilon_{l}(t)
\nonumber\\
&=&\sum_{k=1}^{K}\beta_ke^{-j\frac{2\pi}{\lambda}2d_k(0)}e^{j\frac{2\pi}{\lambda}\eta_l^{r}(\theta_k)}e^{j{2\pi}f_kt}{\bf
x}^T(t){\bf v}(\theta_k)+\epsilon_{l}(t)
\end{eqnarray}
where $\lambda$ is the transmitted signal wavelength, $f_k=2v_kf/c$
is the Doppler shift caused by the $k$-th target, and
\begin{eqnarray}
{\bf
v}(\theta_k)&=&[e^{j\frac{2\pi}{\lambda}\eta^t_1(\theta_k)},...,e^{j\frac{2\pi}{\lambda}\eta^t_{M_t}(\theta_k)}]^T\\\mathrm{and}\
{\bf x}(t)&=&[x_1(t),...,x_{M_t}(t)]^T.
\end{eqnarray}

On letting $L$ denote the number of snapshots and $T_s$ the sampling
period, the received samples collected during the $m$-th pulse are
given by
\begin{eqnarray} \label{8}
{\bf z}_{lm}&=&\left[
                \begin{array}{c}
                  z_l((m-1)T+0T_s) \\
                  \vdots \\
                  z_l((m-1)T+(L-1)T_s) \\
                \end{array}
              \right]=
\sum_{k=1}^{K}\gamma_{k}e^{j\frac{2\pi}{\lambda}\eta_l^{r}(\theta_k)}e^{j{2\pi}f_k(m-1)T}{\bf
D}(f_k){\bf X}{\bf v}(\theta_k)+{\bf e}_{lm}
\end{eqnarray}
where
\begin{eqnarray}
\gamma_{k}&=&\beta_ke^{-j\frac{2\pi}{\lambda}2d_k(0)},\nonumber\\
{\bf
D}(f_k)&=&{\rm diag}\{[e^{j{2\pi}f_k0T_s},\ldots,e^{j{2\pi}f_k(L-1)T_s}]\},\nonumber\\
 {\bf e}_{lm}&=&[\epsilon_{l}((m-1)T+0T_s), \ldots, \epsilon_{l}((m-1)T+(L-1)T_s)]^T,\nonumber\\
\mathrm{and} \ {\bf X}&=&[ {\bf x}(0T_s), \ldots, {\bf
x}((L-1)T_s)]^T \quad (L\times M_t).
\end{eqnarray}

In this paper we assume that the Doppler shift is small, i.e.,
$f_kT_s<<1$ for $k=1,...,K$,  due to slowly moving targets.

\section{Compressive Sensing for MIMO Radar}

Let us discretize the   angle-Doppler plane on a fine grid:
\begin{eqnarray}
\mathbf{a}=[(a_1,b_1),\ldots,(a_N,b_N)].
\end{eqnarray}
We can rewrite (\ref{8}) as
\begin{eqnarray}\label{received signal}
{\bf
z}_{lm}=\sum_{n=1}^{N}s_ne^{j\frac{2\pi}{\lambda}\eta_l^{r}(a_n)}e^{j{2\pi}b_n(m-1)T}{\bf
D}(b_n){{\bf X}}{\bf v}(a_n)+{\bf e}_{lm}
\end{eqnarray}
where
\begin{eqnarray}\label{s_def}
 s_n = \left\{
\begin{array}{rl}
\gamma_{k},  &  \text{if the $k$-th target is  at}\ (a_n,b_n) \\
0,  & \text{otherwise}
\end{array} \right.  \ .
\end{eqnarray}

In matrix form we have
\begin{equation}
{\mathbf z}_{lm}=\mathbf{\Psi}_{lm}{\mathbf{s}} + {\mathbf e}_{lm}
\end{equation}
where $\mathbf{s}=[s_1,\ldots,s_N]^T$ and
\begin{eqnarray}\label{basis_matrix}
\mathbf{\Psi}_{lm}=[e^{j\frac{2\pi}{\lambda}\eta_l^{r}(a_1)}e^{j{2\pi}b_1(m-1)T}{\bf
D}(b_1){{\bf X}}{\bf
v}(a_1),\ldots,e^{j\frac{2\pi}{\lambda}\eta_l^{r}(a_N)}e^{j{2\pi}b_N(m-1)T}{\bf
D}(b_N){{\bf X}}{\bf v}(a_N)].
\end{eqnarray}

Assuming that there are only  a small number of targets, the
positions of targets are  sparse in the angle-Doppler plane, i.e.,
$\mathbf{s}$ is a sparse vector.
 Let us measure linear
projections of ${\bf z}_{lm}$ as
\begin{eqnarray} \label{receive_sig}{\bf
r}_{lm}=\mathbf{\Phi}_{lm}\mathbf{z}_{lm}={\mathbf{\Phi}_{lm}\mathbf{\Psi}_{lm}}\mathbf{s}
+ \tilde {\bf e}_{lm},
\end{eqnarray}
where $\mathbf{\Phi}_{lm}$ is an $M\times L \ (M<L)$ zero-mean
Gaussian random matrix that has small correlation with
${\mathbf{\Psi}}_{lm}$, and  $\tilde {\bf
e}_{lm}=\mathbf{\Phi}_{lm}{\bf e}_{lm}.$  $M$ must be larger than
the number of targets.

If the $l$-th  node in the network knows who the transmit  nodes are
and also knows the transmitters' coordinates relative to a fixed
point in the network, then the node can construct the matrix ${\bf
\Psi}_{lm}$ (\ref{basis_matrix}) and  recover ${\bf s}$ via
$l_1$-optimization based on the node's own received data ${\bf
r}_{lm}$ (see (\ref{receive_sig})). Information on other nodes'
locations could  be provided by higher network layers. If no such
location information is available to the node, or the interference
is strong, then the receive node will pass the linear projections
${\bf r}_{lm}$ to a fusion center, which has global and local
information. Combining the output of $N_p$ pulses at $N_r$ receive
antennas the fusion center can formulate the equation
\begin{eqnarray}\label{cs}
{\bf r}=[{\bf r}^T_{11},\ldots,{\bf r}^T_{1N_p},\ldots,{\bf
r}^T_{N_r1},\ldots,{\bf
r}^T_{N_rN_p}]^T=\mathbf{\Theta}\mathbf{s}+{\bf E}
\end{eqnarray}
where
$\mathbf{\Theta}=[({\mathbf{\Phi}_{11}\mathbf{\Psi}_{11}})^T,\ldots,(
\mathbf{\Phi}_{1(N_p-1)}\mathbf{\Psi}_{1(N_p-1)})^T,\ldots,({\mathbf{\Phi}_{N_r1}\mathbf{\Psi}_{N_r1}})^T,\ldots,(\mathbf{\Phi}_{N_r(N_p-1)}\mathbf{\Psi}_{N_r(N_p-1)})^T]^T$
and ${\bf E}=[ \tilde{\bf e}^T_{11},\ldots, \tilde{\bf
e}^T_{1N_p},\ldots,\tilde{\bf e}^T_{N_r1},\ldots, \tilde{\bf
e}^T_{N_rN_p}]^T$.
Thus, the fusion center can  recover $\mathbf{s}$ by applying the
Dantzig selector to the convex problem of (\ref{cs}) as
(\cite{Candes:07})
\begin{eqnarray}\label{Dantzig}
\hat{{\bf s}}= \min\|{\bf s}\|_1\ \ \ s.t.\
\|{\mathbf{\Theta}}^H({\bf r}-\mathbf{\Theta}{\bf
s})\|_{\infty}<\mu.
\end{eqnarray}
According to \cite{Candes:07},  the sparse vector ${\bf s}$ can be
recovered  with very high probability if $\mu=(1+t^{-1})\sqrt{2\log
N\tilde \sigma^2}\sigma_{max}$, where $t$ is a positive scalar,
$\sigma_{max}$ is the maximum norm of columns in the sensing matrix
$\Theta$ and $\tilde {\sigma^2}$ is the variance of the noise in
(\ref{cs}).  If ${\bf \Phi}{\bf \Phi}^H={\bf I}$ then $\tilde
\sigma^2=\sigma^2$. Determining the best value of $\mu$ requires
some experimentation. A method that requires an exhaustive search
was described  in \cite{Candes:07}. A lower bound  is readily
available, i.e., $\mu>\sqrt{2\log N\tilde \sigma^2}\sigma_{max}$.
Also,
 $\mu$ should not be too large because in that case the trivial solution   ${\bf s}={\bf 0}$ is obtained. Thus,
we may set   $\mu<\|\Theta^H{\bf r}\|_{\infty}$.

\subsection{Resolution}\label{resolution}

 The uniform
uncertainty principle (UUP) \cite{Candes:06}\cite{Candes:08}
 indicates that if every set of columns with cardinality
less than the sparsity of the signal of interest   of the sensing
matrix ($\mathbf{\Theta}$ defined in (\ref{cs})) are approximately
orthogonal, then the sparse signal can be exactly recovered with
high probability. For a fixed $M$   the correlation of columns of
the  sensing matrix can be reduced if the
 number of pulses $N_p$ and/or the number of receive
nodes $N_r$ is increased. Intuitively,  the increase in $N_p$ and
$N_r$ increases the dimension of the sensing matrix columns, thereby
rendering the columns less similar to each other. A more formal
proof is provided in Appendix I.  Moreover, increasing the number of
transmit nodes, i.e., $M_t$, also reduces the correlation of
columns; this is also shown in Appendix I.

In general, to achieve high resolution a fine grid is required.
However, for fixed $N_p$, $N_r$ and $M_t$, decreasing the distance
between the grid points would result in more correlated columns in
the sensing matrix.  Based on the above discussion, the column
correlation can be reduced by increasing $N_p$, $N_r$ or $M_t$.
Also, based on the theory of CS, the effects of a higher column
correlation can be mitigated by using a larger number of
measurements,  i.e., by increasing $M$. In particular, it was shown
in \cite{Candes:06} that $M$ should satisfy $M\geq
\frac{K\epsilon^2(\log N)^4}{C}$, where $\epsilon$ denotes the
maximum mutual coherence between the two columns of the sensing
matrix and $C$ is a positive constant.

One might tend to think that in order to achieve good resolution
one has to involve a lot of measurements, or trasnmit/receive
antennas, or  pulses, which in turn would involve high complexity.
However, extensive simulations suggest that this is not the case. In
fact, the proposed approach can match the  resolution that can
be achieved with conventional methods, while using far fewer
received samples, than those used by the conventional methods.

\subsection{Maximum grid size for the angle-Doppler space }\label{S_maximum_step}

The  grid in the angle-Doppler space must be selected  so that the
targets that do not fall on the chosen grid points can still be
captured by the closest grid points. This requires sufficiently high
correlation of the signal reflected by each target with the columns
of $\mathbf{\Theta}$ corresponding to grid points close to the
targets in the angle-Doppler plane.  However, this requirement goes
against the UUP, {which requires that every set of columns with
cardinality less than the sparsity of the signal of interest  be
approximately orthogonal. Thus, there is a  tradeoff of the
correlation of columns of the sensing matrix and the grid size.}

Absent prior information about the targets, we can determine the
maximum spacing of adjacent grids in the angle-Doppler space by
considering the worst case. Assume that we discretize the
angle-Doppler space uniformly with the spacing $(\Delta a,\Delta b)$
as $\mathbf{a}=[(a_1,b_1),\ldots,(a_N,b_N)]$. The worst case
scenario is that the targets fall in the middle between two adjacent
grid points. Therefore,  a practical approach of selecting the grid
points is to calculate the correlation of columns corresponding to
$(a_i+\frac{\Delta a}{2},b_i+\frac{\Delta b}{2})$ with the columns
corresponding to $(a_i,b_i), \ i=1,\ldots,N$. This can be done by
computing the  correlation at lag zero of columns corresponding to
$(a_i+\frac{\Delta a}{2},b_i+\frac{\Delta b}{2})$ with the columns
corresponding to $(a_i,b_i),$ for $i=1,\ldots,N$, and then taking
 the average.
 Then, we can vary the step
$(\Delta a,\Delta b)$ until the average correlation reaches some
threshold. This threshold should be high enough to capture the
targets that do not fall on the
 grid in the angle-Doppler space, and at the same time, it should   satisfy the UUP. The adoption of such grid points would ensure that the
angle-Doppler estimates of targets  would always fall on the grid of
the constructed basis matrix.

When the targets are between grid points, the increase in $N_p$ or
$N_r$ will not necessarily improve performance. However, simulations
show that we can obtain very good performances with very small $N_p$
and $N_r$. To achieve a similar performance, the conventional
matched filter method will require much greater $N_p$ and $N_r$.

\subsection{Range of unambiguous speed}

Let us assume that the Doppler shift change over the duration $(T)$
of the pulse is negligible as compared to the change between pulses.
This is a reasonable assumption given that we have assumed
$f_kT_s<<1,\ k=1,\ldots, K$. Given two  grid points $(a_i,b_i)$ and
$(a_i,b_j)$ in the angle-Doppler space, where $b_i\neq b_j$, the
corresponding columns of ${\bf \Psi}$ are different if $e^{j2\pi
b_iT}\neq e^{j2\pi b_jT}$. Let $v_i$ be the speed corresponding to
the Doppler frequency $b_i$ and $\Delta_{v}^{ij}=v_j-v_i$. It holds
that
\begin{eqnarray}
e^{j2\pi b_iT}\neq e^{j2\pi b_jT}\ &\Rightarrow&\
\frac{2\Delta_{v}^{ij}fT}{c}\neq n, n=\pm1,\pm2,\ldots
\end{eqnarray}
Therefore, the range of the unambiguous relative speed between two
targets that appear at the same angle satisfies
\begin{eqnarray}
\frac{2\Delta_{v}^{ij}fT}{c}\leq 1\ \Rightarrow \Delta_{v}^{ij} \le
\frac{c}{2fT}.
\end{eqnarray}

The selection of $T$  affects the range of the unambiguous speed;
the smaller  the $T$ the larger the  range of the unambiguous speed
is. We also need a relatively small  $T$ to  satisfy the assumption
that the Doppler shift does not change within the duration of the
pulse. On the other hand, a larger $T$ is needed to satisfy the
narrowband assumption about the transmitted waveforms. Therefore,
$T$ needs to be chosen to balance  the above requirements.

\subsection{Complexity }

The proposed approach requires  solving the convex programming
problem of (\ref{Dantzig}). The more targets one would hope to be
able to detect the  higher the complexity would be. Further, the
signals involved are complex. In this case  (\ref{Dantzig}) can be
recast as a second-order cone program (SOCP) \cite{Candes:05}, which
 requires  polynomial time in the dimension of the unknown
vector.

The requirement of a fine grid further increases the
 computational complexity.  This problem can be mitigated
by first performing an initial angle-Doppler estimation using a
coarse grid, and  then refining the grid points around the  initial
estimate.  Restricting the candidate angle-Doppler space
 reduces the samples in the angle-Doppler space that are required
for constructing the basis matrix, thus reducing the complexity of
the $\ell_1$-optimization step.

In addition to the computation complexity, the receiver for
obtaining the required samples is also more complex. The schematic
diagram of the receiver is shown in Fig. \ref{implementation_scheme}
(see also \cite{Gurbuz:07}).

\section{Performance Analysis in the presence of a jammer signal}
\label{analysis on SJR}
 In \cite{Candes:07},  Candes and Tao showed
that if the basis matrix obeys the UUP and the signal of interest
$\bf s$ is sufficiently sparse, then the square estimation error of
the Dantzig selector satisfies with very high probability
\begin{eqnarray}\label{error_bound}
\parallel\hat{\bf s}-{\bf s}\parallel_{\ell_2}^2 \leq C^2 2log
N\times({\sigma}^2+\sum_{i}^{N}\min(s^2(i),{{\sigma}}^2))
\end{eqnarray}
where C is a constant, $N$ denotes the length of $\bf s$ and
$\sigma^2$ is the variance of the noise. It can be easily seen from
(\ref{error_bound}) that an increase in the interference power
degrades the performance of the Dantzig selector. Thus, in the
presence of  a jammer  that transmits a waveform uncorrelated with
the radar transmit waveforms,  the performance of the proposed CS
method will deteriorate. Next, we provide analytical expressions for
the signal-to-jammer ratio  at the receive nodes, and propose a
modified measurement matrix to suppress the jammer.

\subsection{Analysis of Signal-to-Jammer Ratio}

Suppose that each transmitter transmits $N_p$ pulses. In the
presence of a jammer at location $(d,\theta)$ the signal received at
the $l$-th receive antenna can be expressed as
\begin{eqnarray}
{\bf r}_l= \left[
          \begin{array}{c}
            {\bf r}_{l1} \\
            \vdots \\
            {\bf r}_{lN_p}\\
          \end{array}
        \right]
        &=&\underbrace{\sum_{k=1}^{K}\gamma_ke^{j\frac{2\pi}{\lambda}\eta_l^{r}(\theta_k)}\left[
          \begin{array}{c}
            \mathbf{\Phi}_{l1}e^{j2\pi f_k0T} \\
            \vdots \\
            \mathbf{\Phi}_{lN_p}e^{j2\pi f_k(N_p-1)T}\\
          \end{array}
        \right]{\bf
D}(f_k){\bf X}{\bf v}(\theta_k)}_{{\bf
r}_{ls}}\nonumber\\&&+\underbrace{e^{-j\frac{2\pi}{\lambda}(d-
\eta^r_{l}(\theta))}\beta\left[
          \begin{array}{c}
            \mathbf{\Phi}_{l1}\tilde{{\bf x}}_1\\
            \vdots \\
            \mathbf{\Phi}_{lN_p}\tilde{{\bf x}}_{N_p}\\ \end{array}
        \right]}_{{\bf
r}_{lj}}+\underbrace{\left[
          \begin{array}{c}
            \mathbf{\Phi}_{l1}{\bf e}_{l1}\\
            \vdots \\
            \mathbf{\Phi}_{lN_p}{\bf e}_{lN_p}\\ \end{array}
        \right]}_{{\bf r}_{ln}} \label{23}
\end{eqnarray}
where $\ \tilde{{\bf x}}_{m}=[\tilde{{ x}}_{m}(0T_s),\ldots,\tilde{{
x}}_{m}((L-1)T_s)]^T$ contains  the samples of the signal
transmitted by the jammer during the $m$-th pulse, and $\beta$
denotes the square root of the power of the jammer over the duration
of one signal pulse.

We assume that for all $m$,
$E\{\tilde{x}_m^*(i)\tilde{x}_m(j)\}=1/L$ for $ i=j$, and $0$
otherwise. Thus, $E\{\tilde{{\bf x}}^H_{m} \tilde{{\bf x}}_{m}\}=1$.
Also, we assume that $\tilde{{\bf x}}_m,\ m=0,\ldots,N_p$ are
uncorrelated with the main period of the transmitted waveforms.
 Thus,
the effect of the jammer signal is similar to that of additive
noise. In the following analysis we assume that the jammer
contribution is much stronger than that of additive noise, and
therefore we ignore the third term ${\bf r}_{ln}$ on the  right hand
side of (\ref{23}). Later, in our simulations we will consider
additive noise in addition to a jammer signal.

We assume that all receive nodes use the same random measurement
matrix over $N_p$ pulses, i.e.,
$\mathbf{\Phi}_{l}=\mathbf{\Phi}_{l1}=\mathbf{\Phi}_{l2}=\ldots=\mathbf{\Phi}_{lN_p}$.
Let ${\bf A}_l^{k,k'}={\bf
    X}^H{\bf D}^H(f_k)\mathbf{\Phi}_l^H\mathbf{\Phi}_l{\bf D}(f_{k'}){\bf X}$ and $q^{k,k'}_{i,j}$ denote the $(i, j)$-th element of ${\bf A}^{kk'}_l$.
     Thus, the average power of the desirable signal conditioned on the transmitted waveform
can be represented by
\begin{eqnarray} P_s(l)&=&E\{{\bf r}^H_{ls}{\bf r}_{ls} | {\bf X}\}=
E\{\sum_{k,k'=1}^{K}\underbrace{\gamma_k^*\gamma_{k'}e^{-j\frac{2\pi}{\lambda}(\eta^r_l(\theta_k)-\eta^r_l(\theta_{k'}))}}
_{\rho_{l}(k,{k'})}\underbrace{(\sum_{m=0}^{N_p-1}e^{-j2\pi
(f_k-f_{k'})mT})}_{\mu_{kk'}}\underbrace{{\bf v}^H(\theta_k) {\bf
A}^{kk'}_l {\bf
v}(\theta_{k'})}_{Q_{kk'}}\}\nonumber\\&=&N_p{E\{\sum_{k=1}^{K}|\beta_k|^2
Q_{kk}\}} + E\{\sum_{k\neq k'}\rho_{l}(k,{k'})\mu_{kk'} Q_{kk'}\}
\end{eqnarray}
where $\rho_{l}(k,{k'})$ and $Q_{kk'}$ can be further written as
\begin{eqnarray}
\rho_{l}(k,{k'})&=&e^{j\frac{2\pi}{\lambda}[2(d_k(0)-d_{k'}(0))-(\eta^r_l(\theta_k)-\eta^r_l(\theta_{k'}))}
\beta_k^*\beta_{k'}\\ \mathrm{and} \
 Q_{kk'}&=&\sum_{i,j}q^{k,k'}_{i,j}e^{j\frac{2\pi
}{\lambda}(\eta_j^t(\theta_{k'})-\eta_i^t(\theta_k))}\ .
\end{eqnarray}

 As defined in Section
\ref{sig_model},  the position of the $i$th transmit or receive
(TX/RX) node is denoted by $(r_i^{t/r},\alpha_i^{t/r})$ in polar
coordinates. Thus it holds that
 \begin{eqnarray}
a_{ji}^{k'k}=
{\eta_j^{t/r}(\theta_{k'})-\eta_i^{t/r}(\theta_k)}=\left\{
\begin{array}{rl}
2r^{t/r}_i\sin(\frac{\theta_{k'}-\theta_k}{2})\sin(\alpha_i-\frac{\theta_{k'}+\theta_k}{2})& i=j\\
r^{t/r}_j\cos(\theta_{k'}-\alpha_j)-r^{t/r}_i\cos(\theta_k-\alpha_i)
& i\neq j
\end{array} \right.
 \end{eqnarray}

Let $\psi_0$ be  deterministic. Based on the assumed statistics of
$r_i$ and $\alpha_i$ (see (\ref{2})),   the distribution of
$h=\frac{r_i^{t/r}}{r}\sin(\alpha_i^{t/r}-\psi_0)$ is given by
(\cite{Ochiai})
\begin{eqnarray}
f_h(h)=\frac{2}{\pi}\sqrt{1-h^2},-1<h<1
\end{eqnarray}
and
\begin{eqnarray}\label{property}
E\left\{e^{j\alpha h}\right\}=2\frac{J_1(\alpha)}{\alpha}
\end{eqnarray}
 where
$J_1(\cdot)$ is the first-order Bessel function of the first kind.
Thus, based on (\ref{property}) we can obtain
  \begin{eqnarray}
E\left\{e^{j\frac{2\pi }{\lambda}{a_{ji}^{k'k}}}\right\}
=E\left\{e^{j\frac{2\pi r}{\lambda}\frac{a_{ji}^{k'k}}{r}}\right\}=
\left\{
\begin{array}{rl}
1& i=j\ \text{and}\ k=k'\\
\varsigma(4\sin(\frac{\theta_{k'}-\theta_k}{2}))& i=j\ \text{and}\ k\neq k'\\
\varsigma^2(2) &  i\neq j
\end{array} \right.
 \end{eqnarray}
where $\varsigma(x)=2\frac{J_1(x\frac{\pi r}{\lambda})}{x\frac{\pi
r}{\lambda}}$.

Therefore, the average power of the desirable signal $P_s(l)$ taken
over the positions of TX/RX nodes can be found to be
\begin{eqnarray}\label{signal_power}
P_s(l)&=&N_p{E\left\{\sum_{k=1}^{K}|\beta_k|^2
Q_{kk}\right\}} + E\left\{\sum_{k\neq k'}\rho_{l}(k,{k'})\mu_{kk'} Q_{kk'}\right\}\nonumber\\
&=&N_p\sum_{k=1}^{K}|\beta_k|^2E\left\{ Q_{kk}\right\}+ \sum_{k\neq
k'}E\left\{\rho_{l}(k,{k'})\right\}\mu_{kk'} E\left\{Q_{kk'}\right\}
\nonumber\\
&=&N_p\sum_{k=1}^{K}|\beta_k|^2\sum_{i,j}q^{k,k}_{i,j}E\{e^{j\frac{2\pi
}{\lambda}a_{ji}^{kk}}\}+\sum_{k\neq
k'}\beta_k^*\beta_{k'}e^{j\frac{4\pi}{\lambda}(d_k(0)-d_{k'}(0))}E\{e^{j\frac{2\pi
}{\lambda}a_{ll}^{k'k}}\}\mu_{kk'} \sum_{i,j}q^{k,k'}_{i,j}E\{e^{j\frac{2\pi }{\lambda}a_{ji}^{k'k}}\}\nonumber\\
&= &N_p\sum_{k=1}^K |\beta_k|^2 [\sum_{i}q^{k,k}_{i,i}+ \sum_{i\neq
j}q^{k,k}_{i,j}\varsigma^2(2)] \nonumber\\
&&+\sum_{k\neq
k'}\beta_k^*\beta_{k'}e^{j\frac{4\pi}{\lambda}(d_k(0)-d_{k'}(0))}{\varsigma_{kk'}}\mu_{kk'}
[\varsigma_{kk'}\sum_{i}q^{k,k'}_{i,i}+ \sum_{i\neq j}q^{k,k'}_{i,j}\varsigma^2(2)]\nonumber\\
\end{eqnarray}
where
$\varsigma_{kk'}=\varsigma(4\sin(\frac{\theta_{k'}-\theta_k}{2}))$.

For many practical radar systems with wavelength $\lambda$ less than
$0.1 \mathrm{m}$, (e.g., most military multimode airborne radars),
$2\pi r/\lambda$ is a large number if $r>5m$. Since the function
$\varsigma(x)$ decreases rapidly as $x$ increases,
the terms multiplied by $\varsigma^2(2)$ are  small enough to be
neglected in the above equation. Therefore, (\ref{signal_power}) can
be approximated by
 \begin{eqnarray}
 P_s(l)&\approx&N_p\sum_{k=1}^K |\beta_k|^2 \sum_{i}q^{k,k}_{i,i} +\sum_{k\neq
k'}\beta_k^*\beta_{k'}e^{j\frac{4\pi}{\lambda}(d_k(0)-d_{k'}(0))}{\varsigma_{kk'}}^2\mu_{kk'}
\sum_{i}q^{k,k'}_{i,i}\ .
\end{eqnarray}

Similarly, the average power of the jammer signal over TR/TX
locations is given by
\begin{eqnarray}\label{jammer_power}
P_j(l)&=& E\{{\bf r}_{lj}^H{\bf
r}_{lj}\}=(e^{-j\frac{2\pi}{\lambda}(d-\eta^r_l(\theta))}
\beta)(e^{-j\frac{2\pi}{\lambda}(d-\eta^r_l(\theta))}
\beta)^*\sum_{m=1}^{N_p}\tilde{{\bf
x}}_m^H{\mathbf{\Phi}}^H_l{\mathbf{\Phi}}_l\tilde{{\bf x}}_m\nonumber\\
&=&|\beta|^2\sum_{m=1}^{N_p}\tilde{{\bf
x}}_m^H{\mathbf{\Phi}}^H_l{\mathbf{\Phi}}_l\tilde{{\bf x}}_m.
\end{eqnarray}

The SJR given the node locations is the ratio of the power of the
signal to the power of the jammer. Since the denominator does not
depend on node locations, the average SJR equals
SJR$=P_s(l)/P_j(l)$.

Some insight into the above obtained expression will be given in the
following for some special cases.

\subsection{SJR based on a modified measurement matrix}
Since the jammer signal is uncorrelated with the transmitted signal,
the SJR can be improved by correlating the jammer signal with the
transmitted signal. Therefore, we propose  a measurement matrix of
the form
\begin{eqnarray}\label{measurement_matrix}
\tilde{\mathbf{\Phi}}_l=\mathbf{\Phi}'_l{\bf X}^H \ (M\times L)
\end{eqnarray}
where $\mathbf{\Phi}'_l$ is an $M\times M_t$ Gaussian random matrix.
Note that $\tilde{\mathbf{\Phi}}_l$ is also Gaussian. As stated in
\cite{Candes:08}, a random measurement matrix with i.i.d. entries,
e.g., Gaussian or $\pm1$ random variables, is nearly incoherent with
any fixed basis matrix. Therefore, the proposed measurement matrix
exhibits low coherence with
 $\Psi_l$, thus guaranteeing a stable solution to
(\ref{Dantzig}). Based on (\ref{measurement_matrix}), the average
power of the desirable signal $P_s(l)$ is given by
(\ref{signal_power}),   except that $Q_{kk'}$ is based on ${\bf
A}_l^{k,k'}={\bf X}^H{\bf D}^H(f_k){\bf
X}(\mathbf{\Phi}'_l)^H\mathbf{\Phi}'_l{\bf X}^H{\bf D}(f_{k'}){\bf
X}$. The average power of the jammer signal is given by
(\ref{jammer_power}) where ${\bf \Phi}_l$ is replaced by
$\tilde{{\bf \Phi}}_l$.

Let us assume that the  $M_T$ transmit nodes emit periodic pulses
containing independent quadrature phase shift keying (QPSK) symbols,
and that  ${\bf X}^H{\bf X}={\bf I}_{M_t}$. Also, we assume that
${\mathbf \Phi}_l{\mathbf \Phi}_l^H={\mathbf \Phi}'_l({\mathbf
\Phi}'_l)^H={\bf I}_{M}$.

Let
 $\tilde{x}_i(n)$  be expressed as
$\vartheta_{in}/\sqrt{L}$, where $\vartheta_{in}$ is a random
variable with mean zero and variance one. Then the average power of
the jammer signal $P_j(l)$ can be rewritten as follows:
 \begin{eqnarray}P_j(l)&=& |\beta|^2\sum_{m=1}^{N_p}\tilde{{\bf x}}_m^H{\mathbf{\Phi}}^H_l{\mathbf{\Phi}}_l\tilde{{\bf
x}}_m \nonumber \\
&=&|\beta|^2\sum_{m=1}^{N_p}\sum_{i=j=0}^{L-1}\tilde{{
x}}^*_m(i)\tilde{{
x}}_m(i)c_{ii}+|\beta|^2\sum_{m=1}^{N_p}\sum_{i\neq
j=0}^{L-1}\tilde{{ x}}^*_m(i)\tilde{{ x}}_m(j)c_{ij}\nonumber\\
&=&
\frac{1}{L}|\beta|^2\sum_{m=1}^{N_p}\sum_{i=0}^{L-1}\vartheta^*_{mi}\vartheta_{mi}c_{ii}+\frac{1}{L}|\beta|^2\sum_{m=1}^{N_p}\sum_{i\neq
j=0}^{L-1}\vartheta^*_{mi}\vartheta_{mj}c_{ij}
\end{eqnarray}
where $c_{ij}$ is the $(i, j)$-th entry of
${\mathbf{\Phi}}^H_l{\mathbf{\Phi}}_l$. Since the entries of
${\mathbf{\Phi}}_l$ are i.i.d Gaussian variables with zero means and
variances $\frac{1}{L}$, $c_{ii},i=1,\ldots, L$ are i.i.d chi-square
random variables with means $\frac{M}{L}$ and variances
$\frac{2M}{L}$; $c_{ij},i\neq j$ are of mean zero and variance
$M/L^2$. Let us express $c_{ij}, i\neq j$  as
$\varrho_{ij}\sqrt{M}/L$, where $\varrho_{ij}$ has zero mean and
unit variance.  It holds that
 \begin{eqnarray}
 P_j(l)&=&
 |\beta|^2\sum_{m=1}^{N_p}E\{\vartheta^*_{mi}\vartheta_{mi}c_{ii}\}+\frac{\sqrt{M}}{L^2}|\beta|^2\sum_{m=1}^{N_p}\sum_{i\neq j=0}^{L-1}\vartheta^*_{mi}\vartheta_{mj}\varrho_{ij}\nonumber\\
 &=&|\beta|^2 N_p\frac{M}{L}+\frac{|\beta|^2\sqrt{M}(L-1)}{L}\sum_{m=1}^{N_p}\frac{1}{L(L-1)}\sum_{i\neq j=0}^{L-1}\vartheta^*_{mi}\vartheta_{mj}\varrho_{ij}\nonumber\\
&=&
N_p|\beta|^2\frac{M}{L}+\frac{|\beta|^2\sqrt{M}(L-1)}{L}\sum_{m=1}^{N_p}E\{\vartheta^*_{i,m}\vartheta_{j,m}\varrho_{ij}\}\nonumber\\
&\approx &N_p|\beta|^2\frac{M}{L}
\end{eqnarray}
where we have used the fact that for large $L$,
\begin{eqnarray}
\frac{1}{L} \sum_{i=0}^{L-1}\vartheta^*_{mi}\vartheta_{mi}c_{ii}
&\rightarrow&
E\{\vartheta^*_{mi}\vartheta_{mi}c_{ii}\}=\frac{M}{L}\\ \mathrm{and}
\ \frac{1}{L(L-1)}\sum_{i\neq
j=0}^{L-1}\vartheta^*_{mi}\vartheta_{mj}\varrho_{ij}&\rightarrow&
E\{\vartheta^*_{mi}\vartheta_{mj}\varrho_{ij}\}=0\ .
\end{eqnarray}

Using the  measurement matrix $ \tilde{\mathbf{\Phi}}_l$ in
(\ref{measurement_matrix}) will not affect the average $P_j(l)$ over
the jammer signal due to the fact that
$\sum_{i}c_{ii}=\mathrm{Tr}\{{\bf
X}(\mathbf{\Phi}'_l)^H\mathbf{\Phi}'_l{\bf X}^H\}=\mathrm{Tr}\{{\bf
X}^H{\bf
X}(\mathbf{\Phi}'_l)^H\mathbf{\Phi}'_l\}=\mathrm{Tr}\{\mathbf{\Phi}'_l(\mathbf{\Phi}'_l)^H\}=\mathrm{Tr}\{{\bf
I}_M\}=M$.

In the following, we will look into the SJR improvement using $
\tilde{\mathbf{\Phi}}_l$ as opposed to $ {\mathbf{\Phi}}_l$, for two
different cases, i.e., stationary targets and moving targets.

\subsubsection{ Stationary Targets}
First, let us  consider the SJR using the  random measurement matrix
$\mathbf{\Phi}_l$.

 When the targets are stationary, the Doppler shift
is zero and so $ {\bf A}_l^{k,k'}={\bf A}_l={\bf
    X}^H\mathbf{\Phi}_l^H\mathbf{\Phi}_l{\bf X}
$. Therefore, the average power of the desired signal can be
approximated as
\begin{eqnarray}\label{sig_power1}
 P_s(l)&\approx&N_p\sum_{k=1}^K |\beta_k|^2 \sum_{i}q_{i,i} +N_p\sum_{k\neq
k'}\beta_k^*\beta_{k'}e^{j\frac{4\pi}{\lambda}(d_k(0)-d_{k'}(0))}\varsigma_{kk'}^2
\sum_{i}q_{i,i}
\end{eqnarray}
 where $q_{i,j}$ is the $(i,j)$-th entry of $ {\bf A}_l$.

{Letting ${\bf x}_i$ denote the $i$-th column of ${\bf X}$,
$\sum_{i}q_{i,i}$ can be expressed as}
\begin{eqnarray}
\sum_{i}q_{i,i}&=&\mathrm{Tr}\{{\bf A}_l\}=\sum_{i=1}^{M_t}{\bf
x}^H_i\mathbf{\Phi}_l^H\mathbf{\Phi}_l{\bf
x}_i=\sum_{i=1}^{M_t}\sum_{m,n=1}^Lx^*_i(m)c_{mn}x_i(n)\nonumber\\
&=&\sum_{i=1}^{M_t}\sum_{m=1}^Lx^*_i(m)x_i(m)c_{mm}+\sum_{i=1}^{M_t}\sum_{m\neq
n}^Lx^*_i(m)x_i(n)c_{mn}.
\end{eqnarray}

{The entries of ${\bf X}$ have zero means and mutually independent;
therefore, for sufficiently long $L$ and $M_t$ it holds that
\begin{eqnarray}\label{approx_q}
\sum_{i}q_{i,i}&=& \frac{M_t}{L}\sum_{m=1}^{L}c_{mm}=\frac{M
M_t}{L}.
\end{eqnarray}
Based on (\ref{approx_q}), a  concise form of $P_s(l)$ is given by
\begin{eqnarray}\label{sig_power1}
 P_s(l)&\approx&
\frac{N_pMM_t\sum_{k=1}^K|\beta_k|^2}{L}+\frac{N_pMM_t}{L}\varphi
\end{eqnarray}
 where  $\varphi=\sum_{k,k', k\neq
k'}\beta_k^*\beta_{k'}e^{j\frac{4\pi}{\lambda}(d_k(0)-d_{k'}(0))}\varsigma_{kk'}^2$.

Thus, the SJR corresponding to the random measurement matrix
$\mathbf \Phi_l$ is
\begin{eqnarray}\label{sjr1_no_doppler}
SJR_l=\frac{P_s(l)}{P_j(l)}\approx
\frac{M_t(\sum_{k=1}^K|\beta_k|^2+\varphi)}{|\beta|^2} \ .
\end{eqnarray}

 When using the measurement matrix
$\tilde{\mathbf{\Phi}}_l=\mathbf{\Phi}'_l{\bf X}^H$,  the quantity
corresponding to ${\mathbf A}_l^{k,k'}$ is
\begin{eqnarray}
\tilde{{\bf A}}_l^{k,k'}=\tilde{{\bf A}}_l={\bf
    X}^H{\bf X}(\mathbf{\Phi}'_l)^H\mathbf{\Phi}'_l{\bf X}^H{\bf X}=(\mathbf{\Phi}'_l)^H\mathbf{\Phi}'_l\ .
    \end{eqnarray}

It holds that
$\sum_{i}q_{i,i}=\mathrm{Tr}\{(\mathbf{\Phi}'_l)^H\mathbf{\Phi}'_l\}=\mathrm{Tr}\{\mathbf{\Phi}'_l(\mathbf{\Phi}'_l)^H\}=M$.
Similarly, the average power of the desired signal can be
approximated as
\begin{eqnarray}\label{sig_power2}
 P_s(l)&\approx&N_pM(\sum_{k=1}^K
 |\beta_k|^2+\varphi).
\end{eqnarray}

Therefore, the SJR corresponding to the random measurement matrix
$\tilde{\mathbf{\Phi}}_l$ is
\begin{eqnarray}\label{sjr2_no_doppler}
SJR_l=\frac{P_s(l)}{P_j(l)}\approx
\frac{L(\sum_{k=1}^K|\beta_k|^2+\varphi)}{|\beta|^2}\ .
\end{eqnarray}

From (\ref{sjr1_no_doppler}) and (\ref{sjr2_no_doppler}),  it can be
seen that the use of $\tilde{\mathbf{\Phi}}_l$ instead of
${\mathbf{\Phi}}_l$ can improve SJR by a factor of $L/M_t$
 when $L\gg M_t$. The SJR can be improved by an increase in $L$.
  However, increasing $L$ will require a higher sampling rate when the pulse duration is fixed.
It is interesting to note  that the SJR of (\ref{sjr2_no_doppler})
does not depend on the
  the number of measurements, $M$.

\subsubsection{Slowly Moving Targets}

For simplicity, we consider only the scenarios in which $f_kT<<1$.

Based on the measurement matrix $\mathbf{\Phi}_l$, and considering
the Doppler shift, we have ${\bf A}_l^{k,k'}={\bf
    X}^H{\bf D}^H(f_k)\mathbf{\Phi}_l^H\mathbf{\Phi}_l{\bf D}(f_{k'}){\bf X}$.
   When  the normalized Doppler frequency $f_kT_s\leq 1$,   we have
\begin{eqnarray}\label{Ps_moving}
\sum_{i}q^{k,k'}_{i,i}=\mathrm{Tr}\{{\bf
A}_l^{k,k'}\}=\mathrm{Tr}\{{\bf
    X}^H{\bf D}^H(f_k)\mathbf{\Phi}_l^H\mathbf{\Phi}_l{\bf D}(f_{k'}){\bf
    X}\}\approx \frac{MM_t}{L}.
\end{eqnarray}
Thus,  $P_s(l)$ for the moving targets with $f_sT<<1$ is
approximately the same as that of  stationary targets.

 Let us now consider the measurement matrix $\tilde{ \mathbf{\Phi}}_l$.
Let $c_{ij}^k$ denote the $(i,j)$-th entry of $ {\bf
    X}^H{\bf D}^H(f_k){\bf X}$  and  note that $c_{ij}^k$ is given by $c_{ij}^k=\sum_{n=0}^{L-1}x_i^*(n)x_j(n)*e^{j2\pi
    f_knT_s}$.
In scenarios in which  $f_kT_s<<1$
 and $L$ is relatively large, the
following approximations are readily derived:
\begin{eqnarray}
c_{ij}^k\left\{
\begin{array}{cccc}
&=&\frac{1}{L}\frac{1-e^{j2\pi
    f_kLT_s}}{1-e^{j2\pi
    f_kT_s}}&  i=j\\
&\approx& 0 & i\neq j
\end{array} \right.\ .
\end{eqnarray}
Since the off-diagonal elements are
 small compared with the diagonal elements, they can be ignored.

Then, we  obtain the following approximation
\begin{eqnarray}
{\bf A}_l^{k,k}={\bf
    X}^H{\bf D}^H(f_k){\bf X}(\mathbf{\Phi}'_l)^H\mathbf{\Phi}'_l{\bf X}^H{\bf D}(f_k){\bf
    X}\approx (\mathbf{\Phi}'_l)^H\mathbf{\Phi}'_l.
\end{eqnarray}

Therefore, the SJR of  moving targets with $f_sT<<1$ is
approximately equal to that of stationary targets for both random
measurement matrices.

\section{Simulation Results }\label{simulation}

The goal of this section is to demonstrate the ability of the
proposed MIMO radar approach, denoted in the figures as CS, to pick
up targets in the presence of noise and/or a jammer, and also show
the effect on the various parameters involved. In each case the
performance is compared against other methods that have been
proposed in the context of MIMO radar (here referred to as
``conventional") in order to quantify weaknesses and advantages. For
the case of stationary targets, the conventional methods tested here
are the methods of Capon, APES, GLRT \cite{Xu:06} and MUSIC
\cite{Krim:2006}, while for moving targets, comparison to
  the matched filter method \cite{Levanon:04} is conducted.

In our simulations we consider a MIMO radar system with the
transmit/receive antennas uniformly distributed on a disk of radius
$10$m.  The carrier frequency is $f=5 GHz$ and the sampling rate
$f_s=\frac{1}{T_s}=20M Hz$. The pulse repetition interval is
$T=1/4000 s$.  Each transmit node uses uncorrelated QPSK waveforms.
The received signal is
 corrupted by  zero mean Gaussian noise.  We also consider a jammer that transmits waveforms
  uncorrelated to the signal waveforms.
  For simulation purposes we take the jamming waveforms to be white Gaussian
\cite{Curry:05}. The SNR is defined as the ratio of  power of
transmit waveform  to that of   thermal noise at a receive node.

\subsection{Stationary Targets}

The presence of a target can be seen  in the plot of the magnitude
of  $\hat{{\bf s}}$ obtained by (\ref{Dantzig}).  We will refer to
this vector as {\it target information vector}. The location and magnitude of a peak in that plot
provides target location and RCS  magnitude, respectively.
 The proposed approach results in a clean plot away from the target locations, and well distinguished peaks corresponding to the targets.
 This is a desirable behavior for target detection, as it would result in small
 probability of false alarm. To demonstrate the appearance of the graph we define the  peak-to-ripple ratio (PRR) metric as follows.
 For the $k$-th target,  PRR$_k$ is
 the ratio of the square amplitude of the DOA estimate
at the target azimuth angle to the sum of the square amplitude of
DOA estimates at other angles except at the jammer location, i.e.,
$PRR_k=\frac{|s_k|^2}{{\bf s}^H{\bf
s}-\sum_{i=1}^{K}|s_k|^2-|s_j|^2}$, where ${\bf s}$ is the defined
in (\ref{s_def}), $s_k$ and $s_j$ denote the  elements of $\bf s$
corresponding to  the location of
the $k$-th target and the jammer, respectively.  A clean plot would
yield a high PPR, while a plot with a lot of ripples would yield a
low PRR.

A metric that shows the degree to which a jammer is suppressed,
namely the peak-to-jammer ratio (PJR), is also used here. PJR is defined as the
ratio of the average square amplitude  of the DOA estimates at the target
 angles to the square amplitude of DOA estimates at the
jammer, i.e., $PJR=\frac{\frac{1}{K}\sum_{i=1}^{K}|s_k|^2}{|s_j|^2}$.
Unlike PRR, PJR is averaged over all targets.
In this way, the jammer is considered to be suppressed only  if  the peak amplitude  at
the jammer location is much smaller than the peak amplitude  at any target
location.

The results that we show represent $1,000$ Monte Carlo simulations
over independent waveforms and noise realizations. To better show
the statistical behavior of the methods we plot
 the cumulative density function  (CDF) of PPR and PJR, i.e., $ Probability (PPR < x)$ and $Probability ( PJR < x)$, where $PPR$ is the union of $PRR_k,k=1,\ldots, \ K$.

\subsubsection{Targets falling on the grid}

We consider the following scenario.  Two targets   are located at
angles $\theta_1=0.2\textordmasculine$ and
$\theta_2=-0.2\textordmasculine$. The corresponding reflection
coefficients are $\beta_1=\beta_2=1$.
 A jammer is
located at angle $7 \textordmasculine$ and transmits an unknown
zero-mean Gaussian random waveform with variance $\beta^2=400$. Additive white Gaussian noise is
added at the receive nodes.  The ratio of
the power of transmitted waveforms at each transmit node  to the
variance of the additive Gaussian noise is set to $0$ dB.
The number of transmit antennas is fixed at
 $M_t=30$. For the purpose of reducing computation time, the angle space is taken to be $[-8\textordmasculine, 8\textordmasculine]$, and is sampled with  increments
of $0.2\textordmasculine$ from $-8\textordmasculine$ to
$8\textordmasculine$, i.e., ${\bf
a}=[-8\textordmasculine,-7.8\textordmasculine,\ldots,7.8\textordmasculine,8\textordmasculine]$.
The received signal in a single pulse is sampled, and  $M=30$ random
measurements of one pulse are used to feed the Dantzig selector.
Since the MUSIC method requires  the number of receive antennas to
be greater than the number of targets, when  only one receive
antenna is used we  compare the proposed CS method with only the
Capon, APES and  GLRT methods.  The comparison methods are using
$L=512$ samples to obtain their estimates, while the proposed
approach uses $M=30$ samples. The result of one realization  for the
case of one receive node is shown in Fig.
\ref{PPR_com_five_betaj_20_Mt_30}. One can observe the cleaner
appearance of the graph corresponding to the proposed approach,
 where the two targets appear correctly except with a small error in the magnitude of the target RCS.  The CDF of the corresponding PRR and PJR are also shown in the same figure.
 One can clearly see that with one receive antenna the comparison methods yield PRR close to $1$, which is indicative of severe ripples.

In general, an increase in the number of transmit snapshots $L$
leads to improved PRR and PJR for all methods. In the following
results we fix $L$ to $512$. For the comparison methods, $L$
represents that number of samples needed to obtain target
information. For CS, the number of samples used to extract target
information is $M$.

For the scenario  of Fig. \ref{PPR_com_five_betaj_20_Mt_30},  the
effect of the threshold $\mu$ is evaluated in terms of the empirical
CDF of the PRR and the amplitude estimate of RCS, and the results
are shown   in Fig. \ref{cdf}.  One can can see that the increase in
$\mu$ can lead to fewer ripples but at the same time it degrades the
amplitude estimate of RCS. In the following, the values of $\mu$
used in each case will be shown on the figures.

 For the same target and jammer configuration as above, we now  examine the effect of different levels of jammer strength.
We consider the scenario where $N_r=10$ receive nodes participate in
the estimation. For the case of CS, each node sends to the fusion
center $M=30$ received samples, while for the comparison methods,
each node sends to the fusion center $L=512$ received samples.
 In Fig. \ref{com_jammer} we show the CDF  of PPR and PJR corresponding to jammer variance $\beta^2=400, 1600$ and $3600$ and  SNR equal to $0$ dB.
   One can see that for CS, the probability of low PRR and PJR  increases when the jammer  becomes stronger.
  In particular, there is some non-zero probability that the  PRR will be close to $10^{-7}$. Such cases are rare and  occur when one of
  the two targets is missed. The increase in the threshold $\mu$ can improve the DOA
estimates at the target locations and reduce the probability of
missing one of the targets. The cost, however, would be an increase in ripples.
The performance of the proposed approach can be improved, i.e., the rare low PPR values can be completely avoided by increasing $N_r$,
or $M$. This is demonstrated in Fig. \ref{com_five_betaj_60_all_combination}, where the strong jammer case of Fig. \ref{com_jammer} is considered,
  i.e., $\beta^2=3600$, and $N_r$ is increased to $30$.
We should note here that it does not help to increase $M$ beyond
$M_t$ as the  maximal rank of $\Phi'_l$  is $M_t$.

Next, we consider the same scenario as above but let the two targets
be at variable distance $d$ {in the angle domain}. Figure
\ref{com_spacing} demonstrates
 performances for the cases $d=0.2\textordmasculine, 0.3\textordmasculine, 0.4\textordmasculine$ in the presence
of a strong jammer with variance $\beta^2=3600$. The SNR is $0$ dB,
$N_r=10$ and $M=30$. One can see that the comparison methods produce
good level PRR. Regarding the PJR, as expected, MUSIC fails, Capon
and APES results is PRR$\approx 1$ most of the time, while  GLRT performs well
all the time. The proposed CS approach performs well with a few
exceptions in which a PRR or PJR less than $1$ is obtained with very
small probability.  Again, the CS method performance can be improved by
increasing
 $N_r$ and/or  $M$.

Based on the above results, the performance of the proposed approach
for the jammer dominated scenario can be made at least comparable to
that of the conventional methods while using about $5.8\% \  (=30/512)$ of the number of samples required by the conventional methods.

%

Next, we  study a thermal noise dominated case, i.e., SNR=$-40$dB.
Figure \ref{com_jammer_snr_-40} shows PRR and PJR performance for
different values of jammer variance, i.e., $\beta^2=400,1600$ and
$3600$. In all cases it was taken $N_r=10$,  $M_t=M=30$ and the
targets were separated by $d=0.4\textordmasculine$.  CS yields good
performance even in the presence of both a strong jammer and thermal
noise. The PRR performance of other methods appear to deteriorate at
this noise level. The performances for targets with spacing
$d=0.2\textordmasculine,0.3\textordmasculine$ and
$0.4\textordmasculine$ are given in Fig. \ref{com_spacing_snr_-40}
for $N_r=20$, $M_t=M=30$ and $\beta^2=400$. Like in the case of a
strong jammer, the decrease in the spacing $d$ does not affect the
performance significantly. In this thermal noise dominated case, CS
appears to perform very well in terms of  PRR, and PJR, while the
comparison methods appear to be very noisy. To further examine this
case, we consider two additional performance measures, i.e.,  mean
squared error (MSE)  and probability of false alarm (PFA), which are
computed  based on the obtained estimate $\hat {\bf s}$ as follows.
A new vector, $\hat {\bf s}^t$ is formed; if $\hat {s}_i$ is greater
than some threshold then $\hat { s}^t_i=1$, otherwise, $\hat {
s}^t_i=0$.
 The MSE  is calculated as
$MSE=\|\hat{{\bf s}}_t-{\bf s}_t\|^2_2/N$, where ${\bf s}_t$ is an
$N\times 1$ vector which contains zeros everywhere except   at angles corresponding to  target locations, where it is $1$.
The PFA measures the probability of $1$ occurring in $\hat{{\bf s}}$
at non-target locations. Figure \ref{mse_pfa} shows the MSE  based on $8,000$ Monte Carlo simulations. Note that the performance of
MUSIC is not shown here since MUSIC always yields a peak at the jammer
location. One can see that the simple thresholding described above  helps the comparison methods,  and if
the threshold is picked appropriately all methods can produce a low
angle MSE and PFA.
 However, the
 MSE corresponding to the CS method is less sensitive to the particular
threshold than other methods. For the milder jammer case
($\beta=20$), the CS approach exhibits slightly better ``best MSE
performance" than the comparison methods, while in the stronger
jammer case ($\beta=60$)  GLRT outperforms CS for most thresholds.
For the strong jammer case, the MSE and PFA of  CS are compared to
those  of GLRT for different number of samples, $L$ in Fig.
\ref{mse_pfa_new}. One can see that  for the strong jammer case
($\beta=60$) CS performs comparably to GLRT with $L=256$. Thus, in
the strong jammer case,  CS still achieves good performance with
fewer samples than GLRT, except that the savings in terms of number
of samples is smaller. For CS, the trend of an increasing MSE as the
threshold increases can be explained by the fact that one of the two
targets can be missed as the threshold increases. GLRT relies on the
Gaussian assumption for the noise and jammer signals, which is
totally valid in our simulations. Thus, unlike the other methods,
GLRT can suppress the jammer completely. We should note that the
specific values of MSE and PFA depend on the kind of thresholding
performed. For example, applying thresholding on a  nonlinear
transformation of the estimated vector can give different MSE and
PFA, and the best results for each method are not necessarily
obtained based on the same non-linear transformation. Determining
the best thresholding method is outside the scope of this paper.

\subsubsection{Targets falling off the grid points}
In this section, we consider  scenarios in which targets do not fall
on the grid points. This is a case of practical interest, as the
target locations are unknown, thus  the best grid in not known in
advance. We first select the proper step to discretize the angle
space following the procedures described in Section
\ref{S_maximum_step}.
The angle space is sampled by increments of $0.2\textordmasculine$
from $-8\textordmasculine$ to $8\textordmasculine$, i.e., ${\bf
a}=[-8\textordmasculine,-7.8\textordmasculine,\ldots,7.8\textordmasculine,8\textordmasculine]$.
Assume that four targets  of interest  are located at
$\theta_k=\{-1.1\textordmasculine,-0.3\textordmasculine,
0.3\textordmasculine,1.1\textordmasculine\}$. Their reflection
coefficients are $\{\beta_k=1, k=1,2,3,4\}$. A jammer is still
located at $7\textordmasculine$. Since the targets are located
between the grid points, we cannot plot PRR and PJR as in the case
of targets onto the grid points. Therefore,  we show  the mean plus
and minus one standard deviation (std) for the amplitude of DOA
estimate at each grid point. The results are shown in Fig.
\ref{com_angle_off_grid}. The power of the jammer  was $400$ (left
column of Fig. \ref{com_angle_off_grid})  and $3600$ (right column).
Based on Fig. \ref{com_angle_off_grid}, it can be seen that with the
proper grid points, the proposed method can capture well the targets
that do not fall on grid points. The next best method is the GLRT
which captures the targets but exhibits high variance as indicated
by the shaded region around the mean.

\subsection{Moving Targets}
We continue to consider  orthogonal QPSK waveforms and a jammer
located at $7\textordmasculine$ with the power $400$. The SNR is
still set to be $0$ dB  and each receive node collects $M=30$
measurements. Figures \ref{angle_doppler_on_grid} and
\ref{angle_doppler_off_grid} show the target scene of the proposed
CS method and {the matched filter approach \cite{Levanon:04} for
targets on the grid points and off the grid points, respectively.
The matched filter correlates  the receive signal  with the transmit
signal distorted by different Doppler shifts and  steering vectors.

\subsubsection{Targets falling onto the grid points}

We assume the presence of three targets located at
$\{\theta_k=-1\textordmasculine,\ 0\textordmasculine,\
1\textordmasculine\}$  that are moving at the speed  of
$\{v_k=60m/s,\ 70m/s,\ 80 m/s\}$, respectively. We sample the
angle-Doppler space by the increment $(0.5\textordmasculine,5m)$ as
\begin{eqnarray}{\bf
a}=[(-8\textordmasculine,50m/s),(-7.5\textordmasculine,50m/s),\ldots,(8\textordmasculine,50m/s),(-8\textordmasculine,55m/s)\ldots,(8\textordmasculine,55m/s),\ldots,\ldots,(8\textordmasculine,110m/s)]
\end{eqnarray}

 Figure \ref{angle_doppler_on_grid} shows the target scene for one realization  corresponding to $N_1=1$ receive nodes (left column of the figure), and also $N_r=10$ (right column of the figure).
  We can
see that the performance of the match filtering method is  inferior to
that of the CS approach even when using the data of $30$ pulses.  The proposed CS
approach can yield the desired performances even with a single
receive node and as low as  $5$  pulses.
Comparing the left column and right column of Fig.
\ref{angle_doppler_on_grid}, one can see the effect of the number
of receive antennas $N_{r}$.  The increase in $N_r$ can reduce the
number of pulses required to produce good performance.

\subsubsection{Targets falling off the grid points}
In this section, we consider the scenarios in which targets that do
not fall on grid points.  
 From simulations (the corresponding figure is not given here because of space limitations), we found that the column correlation  is more
 sensitive to the angle step than the  speed step, since $fT_s<<1$. This
 indicates that in the initial estimation, the grid points should be closely spaced  in the angle axis
and  relatively sparser in the speed axis. Then the resolution of
target detection can be improved by taking  denser samples of the
angle-Doppler space around the initial angle-Doppler estimate.

Like the scenarios with the stationary targets, the angle dimension
is sampled by increments of $0.2\textordmasculine$ and the step of
the
speed dimension is set to  $5 m/s$. 
Three targets are moving at the speed  of $\{v_k=62.5m/s,\ 72.5m/s,\
82.5 m/s\}$ in the direction of $\{\theta_k=-1.1\textordmasculine,\
0.1\textordmasculine,\ 1.1\textordmasculine\}$.
Fig.\ref{angle_doppler_off_grid} demonstrates that the proposed
method can capture the targets which fall out of the grid points in
both angle and speed dimensions and it can outperform the
conventional matched filter method. Moreover, we can see that    the
increase in $N_p$ or $N_r$ will not necessarily improve  performance
for the targets between  grid points. This is because an increase in
the dimension of the basis vectors will decrease the correlation of
columns in the basis matrix, which contradicts the requirement for
capturing the targets out of the grid points \ref{S_maximum_step}.
The performance in the case of closer spaced targets, i.e., $d=0.4\textordmasculine$ is shown in Fig. \ref{angle_doppler_on_grid_snr_0_d_0.4}.

\section{Conclusions}
We have proposed a  MIMO radar system that can be implemented by a
small-sized
 wireless network. Network nodes serve as transmitters or receivers.
 Transmit nodes transmit uncorrelated waveforms. Each receive node applies compressive sampling to the received signal to obtain a small number of samples, which the node subsequently forwards to a
  fusion center.
  Assuming that the targets are sparsely located in the
angle-Doppler space,  the fusion center formulates an
$\ell_1$-optimization problem,  the solution of which yields target
angle and Doppler information. For the stationary case, the
performance of the proposed approach was compared to that of
conventional approaches that have been proposed in the context of
MIMO radar. The comparison scenario assumed that each receive node
forwards the received signal to a fusion center, where Capon, APES,
GLRT or MUSIC is implemented to obtain target information. The proposed approach can extract target
information based on a small number of measurements from one
of more receive nodes.
 In particular, for a mild jammer, the proposed method  has been shown to be at least as good  as   the
Capon, APES, GLRT and MUSIC techniques while using a significantly smaller number of samples.
In the case of strong thermal noise and strong
jammer, the proposed method performs slightly worse than the GLRT
method. In that case,  its performance   is still acceptable,
especially if one takes into account the fact that it uses significantly fewer samples than GLRT. For the case of
moving targets, the proposed approach was compared to
 conventional matched filtering, and was shown to perform better  in
both  single and multiple receive nodes cases. An important feature
of the proposed approach is energy savings. If the fusion center
implemented the proposed CS approach, it would require  nodes  to
forward  $M$ samples each, as opposed to $L$ samples that would be
needed if the fusion center implemented the conventional methods. In
order to meet a certain performance, $M$ is typically significantly
smaller than $L$, i.e.,  fewer samples would be needed for the CS
implementation as compared to the implementation of conventional
methods. This translates to energy savings during the transmission
of the samples from the receive nodes to the fusion center. The
obtained savings would be significant in prolonging the life of the
wireless network. Future work includes extension to extracting range
information, and also studying
 scenarios of widely separated antennas and wideband
radar signals. The proposed approach assumes that nodes are
synchronized and the fusion center has perfect node location
information.
 The effects of localization and
synchronization errors and ways to mitigate them need to be further
studied.

\bigskip
\centerline{\bf Acknowledgment} The authors would like to thank Dr.
Rabinder Madan of the Office of Naval Research for sharing his ideas on the use of compressive sampling in the context of MIMO radar, and also the Associate Editor and the anonymous reviewers for their helpful comments.

\bibliographystyle{IEEE}

\begin{thebibliography}{xx}

\bibitem{Fishler:04} E. Fishler, A. Haimovich, R. Blum, D. Chizhik, L. Cimini and R.
Valenzuela,  \newblock `` MIMO radar: An idea whose time has come,"
in \newblock {\em   Proc. IEEE Radar Conf.}, Philadelphia, PA, pp.
71-78, Apr. 2004.


\bibitem{Xu:06} L. Xu, J. Li and P. Stoica, \newblock `` Radar imaging via adaptive MIMO techniques,"  in \newblock {\em Proc. European Signal Process. Conf.}, Florence, Italy, Sep. 2006.



\bibitem{Li:07} J. Li, P. Stoica, L. Xu and W. Roberts, \newblock ``On parameter identifiability of MIMO radar,"
\newblock {\em IEEE Signal Process. Lett.}, vol. 14,  no. 12, pp. 968 - 971,  Dec.
2007.




\bibitem{Haimovich: 08} A. M. Haimovich,  R.S. Blum and L.J. Cimini,  \newblock ``MIMO
radar with widely separated antennas,"
\newblock {\em IEEE Signal
Processing Magazine}, vol. 25, issue 1,  pp. 116 - 129,    2008.


\bibitem{Stoica: 07m} P. Stoica and    J. Li ,  \newblock ``MIMO radar with
colocated antennas,"
\newblock {\em IEEE Signal
Processing Magazine}, vol. 24,   pp. 106 - 114,  issue 5,  2007.

\bibitem{Chen:081} C. Chen and P. P. Vaidyanathan, \newblock ``MIMO radar space-time adaptive processing
using prolate spheroidal wave functions,"
\newblock {\em IEEE Trans. Signal Process.}, vol. 56,  no. 2, pp. 623 - 635,  Feb.
2008.

\bibitem{Stoica:07}P. Stoica,    J. Li   and Y. Xie, \newblock ``On probing signal
design for MIMO radar,"  \newblock {\em  IEEE Trans. Signal
Process.},   vol. 55, pp. 4151-4161, Aug. 2007.

\bibitem{Aittomaki:07}T. Aittomaki and V. Koivunen, \newblock ``Signal covariance matrix
optimization for transmit beamforming in MIMO radars," in
\newblock {\em Proc. 38th Asilomar Conf. Signals, Syst. Comput.},  pp. 182-186, Pacific Grove, CA, Nov. 2007

\bibitem{Fuhrmann:08}D. R. Fuhrmann and G. San Antonio, \newblock ``Transmit beamforming for MIMO
radar systems using signal cross-correlation,"  \newblock {\em  IEEE
Trans.
 Aerospace and Electronic Systems}, vol. 44, pp. 171 - 186, January 2008.

\bibitem{Donoho:06} D. V. Donoho,   \newblock ``Compressed sensing,"
\newblock {\em IEEE Trans. Information Theory}, vol. 52, pp.
1289-1306, no. 4, April 2006.

\bibitem{Candes:06} E. J. Candes, \newblock ``Compressive sampling," \newblock {\em Proceedings of the International Congress of Mathematicians},  Madrid, Spain, 2006.

\bibitem{Candes:08} E. J. Candes and  M. B. Wakin,  \newblock ``An
introduction to compressive sampling [A sensing/sampling paradigm
that goes against the common knowledge in data acquisition],"
\newblock {\em IEEE Signal
Processing Magazine}, vol. 25, pp. 21 - 30 , March 2008.

\bibitem{Candes:05} E. J. Candes and  J. Romberg,  \newblock ``$\ell_1$-MAGIC: Recovery of sparse signals via convex programing,"
\newblock {\em http://www.acm.caltech.edu/l1magic/}, October 2008.

\bibitem{Romberg:08} J. Romberg, \newblock ``Imaging via compressive sampling [Introduction to
compressive sampling and recovery via convex programming],"
\newblock {\em IEEE Signal
Process. Mag.}, vol. 25, no. 2, pp. 14 - 20, Mar. 2008.







\bibitem{Bajwa:06} W. Bajwa,   J. Haupt, A. Sayeed and  R. Nowak, \newblock ``Compressive wireless sensing," in
 \newblock {\em Proc. IEEE Inform. Process. in Sensor Networks},
  Nashville, TN, pp. 134 - 142, Apr. 2006.

 \bibitem{Paredes:07} J. L.Paredes,   G. R. Arce, Z. Wang, \newblock ``Ultra-wideband
  compressed sensing: Channel
estimation,"
 \newblock {\em IEEE Journal of Selected Topics in Signal Processing},   vol.1,  pp. 383 - 395, Oct. 2007.

\bibitem{Baraniuk:07} R. Baraniuk and P. Steeghs, \newblock ``Compressive Radar Imaging,"  \newblock {\em Proc. Radar
Conference},   pp. 128 - 133, April, 2007.

\bibitem{Gurbuz:07} A. C. Gurbuz, J. H. McClellan and W.R. Scott,  \newblock ``Compressive sensing
for GPR imaging,"
\newblock {\em  Proc. 41th Asilomar Conf. Signals, Syst. Comput},   pp. 2223-2227, Pacific Grove, CA, Nov. 2007.


\bibitem{Herman:08}  M. Herman and T. Strohmer, \newblock ``Compressed rensing radar," in \newblock {\em Proc. IEEE Int'l Conf.  Acoust. Speech Signal Process},
Las Vegas, NV, pp. 2617 - 2620, Mar. - Apr. 2008.



\bibitem{Gurbuz:08} A. C. Gurbuz, J. H. McClellan,  V. Cevher, \newblock ``A compressive beamforming method," in
\newblock {\em Proc. IEEE Int'l Conf.  Acoust. Speech Signal Process},  Las Vegas, NV, pp. 2617 - 2620, Mar. - Apr. 2008.

\bibitem{Levanon:04} Nadav Levanon and Eli Mozeson, \emph{Radar
Signals},
 Hoboken, NJ: J. Wiley, 2004.

\bibitem{Petropulu:08} A. P. Petropulu, Y. Yu and H. V. Poor,  \newblock ``Distributed MIMO radar
using compressive sampling,"
\newblock {\em  Proc. 42nd Asilomar Conf. Signals, Syst. Comput}, Pacific Grove, CA,  Nov.
2008.

\bibitem{Chen:08} C. Y. Chen and P. P. Vaidyanathan,  \newblock ``Compressed sensing in MIMO radar,"
\newblock {\em  Proc. 42nd Asilomar Conf. Signals, Syst. Comput},  Pacific Grove, CA, Nov.
2008.

\bibitem{Candes:07} E. Candes and T. Tao, \newblock ``The Dantzig selector: Statistical estimation
when $p$ is much larger than $n$,"  \newblock {\em  Ann. Statist.},
vol. 35, pp. 2313-2351, 2007.

\bibitem{Herman}
M. A. Herman and T. Strohmer, ``High-resolution radar via compressed
sensing,'' to appear in {\it IEEE Trans. Signal Process.}.

\bibitem{Ochiai}
H. Ochiai, P. Mitran, H. V. Poor and V. Tarokh, ``Collaborative
beamforming for distributed wireless ad hoc sensor networks,'' {\it
IEEE Trans. Signal Process.}, vol. 53,  no. 11, pp. 4110 - 4124,
Nov. 2005.

\bibitem{Krim:2006}
H. Krim and M. Viberg, ``Two decades of array signal processing
research: The parametric approach,'' {\it IEEE Signal Processing
Magazine}, vol. 13,   pp. 67 - 94, July 1996.


\bibitem{Curry:05}
G. R. Curry, \emph{Radar system performance modeling},  Boston:
Artech House, 2005.

%





\end{thebibliography}

\begin{biography}
{Yao Yu}  received the B.S. degree in the Telecommunication
Engineering from Xidian University, Xian, China, in 2005, and the
Mphil. degree from City University of Hong Kong, Hong Kong, in 2007.
She is currently a Ph.D. candidate in the Electronic Engineering at
Drexel University, Philadelphia. Her research interests currently
include wireless communication and digital signal processing.
\end{biography}
\begin{biography}
{Athina P. Petropulu} received the Diploma in Electrical Engineering
from the National Technical University of Athens, Greece in 1986,
the M.Sc. degree in Electrical and Computer Engineering in 1988 and
the Ph.D. degree in Electrical and Computer Engineering in 1991,
both from Northeastern University, Boston, MA.

In 1992, she joined the Department of Electrical and Computer
Engineering at Drexel University where she is now a Professor.
During the academic year 1999/2000 she was an Associate Professor at
Universit$\acute{\mathrm{e}}$ Paris Sud, $\dot{\mathrm{E}}$cole
Sup$\acute{\mathrm{e}}$rieure
d'Electr$\acute{\mathrm{i}}$cit$\acute{\mathrm{e}}$ in France. Dr.
Petropulu's research interests span the area of statistical signal
processing, wireless communications and networking and ultrasound
imaging. She is the recipient of the 1995 Presidential Faculty
Fellow Award in Electrical Engineering given by NSF and the White
House. She is the co-author (with C.L. Nikias)  of the textbook
entitled, "Higher-Order Spectra Analysis: A Nonlinear Signal
Processing Framework," (Englewood Cliffs, NJ: Prentice-Hall, Inc.,
1993).

She has served as an Associate Editor for the IEEE Transactions on
Signal Processing and the IEEE Signal Processing Letters, and is a
member of the editorial board of the IEEE Signal Processing Magazine
and the EURASIP Journal on Wireless Communications and Networking.
She is IEEE SPS Vice President-Conferences, member of the IEEE
Signal Processing Board of Governors, member of the IEEE Signal
Processing Society Conference Board and the Technical Committee on
Signal Processing Theory and Methods. She was the General Chair of
the 2005 International Conference on Acoustics Speech and Signal
Processing (ICASSP-05), Philadelphia PA.
\end{biography}
\begin{biography}{H. Vincent Poor }
(S'72, M'77, SM'82, F'87) received the Ph.D. degree in EECS from
Princeton University in 1977.  From 1977 until 1990, he was on the
faculty of the University of Illinois at Urbana-Champaign. Since
1990 he has been on the faculty at Princeton, where he is the
Michael Henry Strater University Professor of Electrical Engineering
and Dean of the School of Engineering and Applied Science. Dr.
Poor's research interests are in the areas of stochastic analysis,
statistical signal processing and their applications in wireless
networks and related fields. Among his publications in these areas
are the recent books MIMO Wireless Communications (Cambridge
University Press, 2007) and Quickest Detection (Cambridge University
Press, 2009).

Dr. Poor is a member of the National Academy of Engineering, a
Fellow of the American Academy of Arts and Sciences, and an
International Fellow of the Royal Academy of Engineering (U.K.). He
is also a Fellow of the Institute of Mathematical Statistics, the
Optical Society of America, and other organizations.  In 1990, he
served as President of the IEEE Information Theory Society, and in
2004-07 he served as the Editor-in-Chief of the IEEE Transactions on
Information Theory. He was the recipient of the 2005 IEEE Education
Medal. Recent recognition of his work includes the 2007 IEEE Marconi
Prize Paper Award, the 2007 Technical Achievement Award of the IEEE
Signal Processing Society, and the 2008 Aaron D. Wyner Award of the
IEEE Information Theory Society.

\end{biography}

\appendices
\section{The effects of $N_r,N_p,M_t$ on the correlation of columns
in the sensing matrix}\label{appendix}
\subsection{The effect of the number of pulses on the column correlation in the sensing matrix}
The sensing matrix for the $l$-th receive antenna
$\mathbf{\Theta}_l$ is given by
\begin{eqnarray}
\mathbf{\Theta}_l=\left[
          \begin{array}{c}
           \mathbf{\Phi}_{l}\mathbf{\Psi}_{l1}  \\
            \vdots \\
           \mathbf{\Phi}_{l}\mathbf{\Psi}_{l(N_p-1)}\\
          \end{array}
        \right]
\end{eqnarray}
where $\mathbf{\Psi}_{lm},m=0,\ldots,N_p-1 $, is defined in
(\ref{basis_matrix}).

On letting ${\bf g}_k$ denote the $i$-th column of
$\mathbf{\Theta}_l$, the correlation of columns ${\bf g}_k$ and
${\bf g}_{k'}$ equals
\begin{eqnarray}\label{correlation}
p_{kk'}&=&|<{\bf g}_k,{\bf g}_{k'}>|
=\left\{
\begin{array}{rl}
N_p|{\bf v}^H(a_k) {\bf B}^{kk}_l {\bf
v}(a_{k})|& k=k'\\
|\frac{sin(\pi (b_k-b_{k'})N_pT)}{\sin(\pi (b_k-b_{k'})T)}||{\bf
v}^H(a_k) {\bf B}^{kk'}_l {\bf v}(a_{k'})|& k\neq k'
\end{array} \right.\ .
\end{eqnarray}
where ${\bf B}^{kk'}_l={\bf
    X}^H{\bf D}^H(b_k)\Phi_l^H\Phi_l{\bf D}(b_{k'}){\bf
    X}$.

 For a
given pair $(k,k'), k\neq k'$, the ratio of $|<{\bf g}_k,{\bf
g}_{k}>|$ to  $|<{\bf g}_k,{\bf g}_{k'}>|$, i.e., $h_{k k'}$,
reveals the effect of
 $N_p$ on the correlation of the two columns. It  holds that
\begin{eqnarray}
h_{kk'} \propto \frac{N_p}{|\sin(\pi (b_k-b_{k'})N_pT)|}\ .
\end{eqnarray}
Let assume that $T$ has been fixed. As long as $(b_k-b_{k'})N_pT\leq
1$, $h_{kk'}$  increases with  $N_p$, and attains the maximum value
 when
$(b_k-b_{k'})N_pT=1$, because the cross correlation of  ${\bf g}_k$
and ${\bf g}_{k'}$ becomes zero. Therefore,  the increase in $N_p$
can improve the performance of CS estimation of (\ref{Dantzig}) as
long as $(b_k-b_{k'})N_pT\leq 1$. This indicates that if
$(b_k-b_{k'})N_pT\leq 1$ for each pair of $(k,k'),k\neq k'$, the
increase in $N_p$ can always improve the performances of CS
estimation. For a conventional radar, the number of pulses can also
improve the resolution of Doppler estimates since  the Doppler shift
creates greater change between pulses.
\subsection{The effect of the number of receive antennas on the column correlation in the sensing matrix}
Next, we  investigate the effect of the number of receive antennas
$N_r$ on the correlation of columns in the sensing matrix. For
simplicity, we assume only the received data collected during the
$n$-th pulse is considered and the random measurement matrix $\Phi$
is constant over receive antennas. Then the sensing matrix $\Theta$
can be represented as
\begin{eqnarray}
\Theta=\left[
          \begin{array}{c}
           \mathbf{\Phi}\mathbf{\Psi}_{1n}  \\
            \vdots \\
           \mathbf{\Phi}\mathbf{\Psi}_{N_rn}\\
          \end{array}
        \right].
\end{eqnarray}

Thus, the correlation of columns ${\bf g}_i$ and ${\bf g}_{j}$
equals
\begin{eqnarray}
p_{ij}&=& |<{\bf g}_i,{\bf g}_{j}>| =\left|\sum_{l=1}^{N_r}
e^{j\frac{2\pi}{\lambda}(\eta_{l}^{r}(a_j)-\eta_{l}^{r}(a_i))}\right|
\left|e^{j2\pi(n-1)T(b_j-b_i)}{\bf
v}^H(a_i){\bf X}^H{\bf D}^H(b_i)\Phi^H\Phi{\bf D}(b_j){\bf X}{\bf
v}(a_j)\right|\nonumber\\
&=&\left\{
\begin{array}{rl}
N_r|{\bf v}^H(a_i){\bf B}^{i,j}{\bf
v}(a_j)|& i=j\\
|\sum_{l=1}^{N_r}
e^{j\frac{2\pi}{\lambda}(\eta_{l}^{r}(a_j)-\eta_{l}^{r}(a_i))}||{\bf
v}^H(a_i){\bf B}^{i,j}{\bf v}(a_j)|& i\neq j
\end{array} \right.\
\end{eqnarray}
where ${\bf B}^{i,j}={\bf X}^H{\bf D}^H(b_i)\Phi^H\Phi{\bf
D}(b_j){\bf X} $.

 Then the ratio of $|<{\bf g}_i,{\bf g}_{j}>|$ to  $|<{\bf
g}_i,{\bf g}_{i}>|$ is
\begin{eqnarray} \label{cor_dif_nr}
h_{ij}\propto \frac{1}{N_r}\left|\sum_{l=1}^{N_r}
e^{j\frac{2\pi}{\lambda}(\eta_{l}^{r}(a_j)-\eta_{l}^{r}(a_i))}\right|.
\end{eqnarray}

 Since the receive nodes are randomly and independently
distributed, $\frac{1}{N_r}|\sum_{l=1}^{N_r}
e^{j\frac{2\pi}{\lambda}(\eta_{l}^{r}(a_j)-\eta_{l}^{r}(a_i))}|$
 approaches  $0$ as $N_r$ becomes large. Therefore,
the correlation of two columns in the sensing matrix can be reduced
when the number of receive antennas is increased.


\subsection{The effect of the number of transmit antennas on the column correlation in the sensing matrix}
Finally, let us see the effect of the number of transmit nodes on
the correlation of columns. For simplicity, we assume  $N_r=N_p=1$.
Then ${\bf v}^H(a_i){\bf B}^{i,j}{\bf v}(a_j)$ can be rewritten as
\begin{eqnarray}
{\bf v}^H(a_i){\bf B}^{i,j}{\bf
v}(a_j)&=&\sum_{k,p}v_k(a_j)v^*_k(a_i)B^{i,j}_{p,p}/L+\underbrace{\sum_{k}\sum_{p\neq
q}v_k(a_j)v^*_k(a_i)x_k(q)x^*_{k}(p)B^{i,j}_{p,q}}_{\sigma_1^{ij}}\nonumber\\
&&+\underbrace{\sum_{k\neq
k'}\sum_{p,q}v_k(a_j)v^*_{k'}(a_i)x_k(q)x^*_{k'}(p)B^{i,j}_{p,q}}_{\sigma_
2^{ij}}\end{eqnarray} \begin{eqnarray} \approx \left\{
\begin{array}{rl}
\frac{MM_t}{L}+\sigma_1^{ii}+\sigma_2^{ii}& i=j\\
\frac{M\sum_kv_k(a_j)v^*_k(a_i)}{L}+\sigma_1^{ij}+\sigma_2^{ij}&
i\neq j
\end{array} \right.\
\end{eqnarray}
where $v_k$ and $X_{p,q}$ denote the $k$-th entry of $\bf v$ and the
$(p,q)$-th entry of $\bf X$, respectively.

Thus,  the ratio of $|<{\bf g}_i,{\bf g}_{j}>|$ to $|<{\bf g}_i,{\bf
g}_{i}>|$ is
\begin{eqnarray} \label{cor_dif_mt}
h_{ij}=\left|\frac{\frac{M\sum_kv_k(a_j)v^*_k(a_i)}{L}+\sigma_1^{ij}+\sigma_2^{ij}}{\frac{MM_t}{L}+\sigma_1^{ii}+\sigma_2^{ii}}\right|
=\left|\frac{\frac{M\sum_kv_k(a_j)v^*_k(a_i)}{M_tL}+\frac{\sigma_1^{ij}}{M_t}+\frac{\sigma_2^{ij}}{M_t}}{\frac{M}{L}+\frac{\sigma_1^{ii}}{M_t}+\frac{\sigma_2^{ii}}{M_t}}\right|.
\end{eqnarray}
It can easily be seen that the numerator approaches 0 as $M_t$
approaches infinity. Therefore, the correlation of two columns of
the sensing matrix can be reduced by  employing a large number of
transmit nodes $M_t$.  

\newpage



\begin{figure}
  \begin{center}
    \scalebox{0.4}{\includegraphics{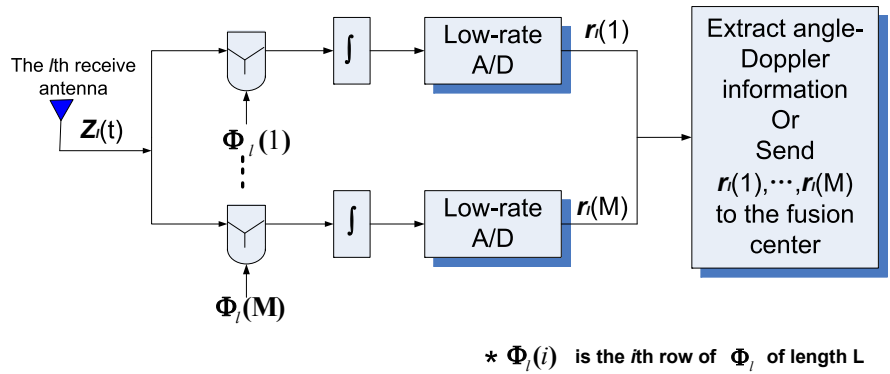}}\end{center}
 \caption{Schematic diagram of the receiver. $\Phi_l $ denotes
 the measurement matrix for the $l$th receive node.
  }\label{implementation_scheme}
\end{figure}

\begin{figure}[htbp]
  \centering
    \includegraphics[height=3.5in,width=5in,clip=true]{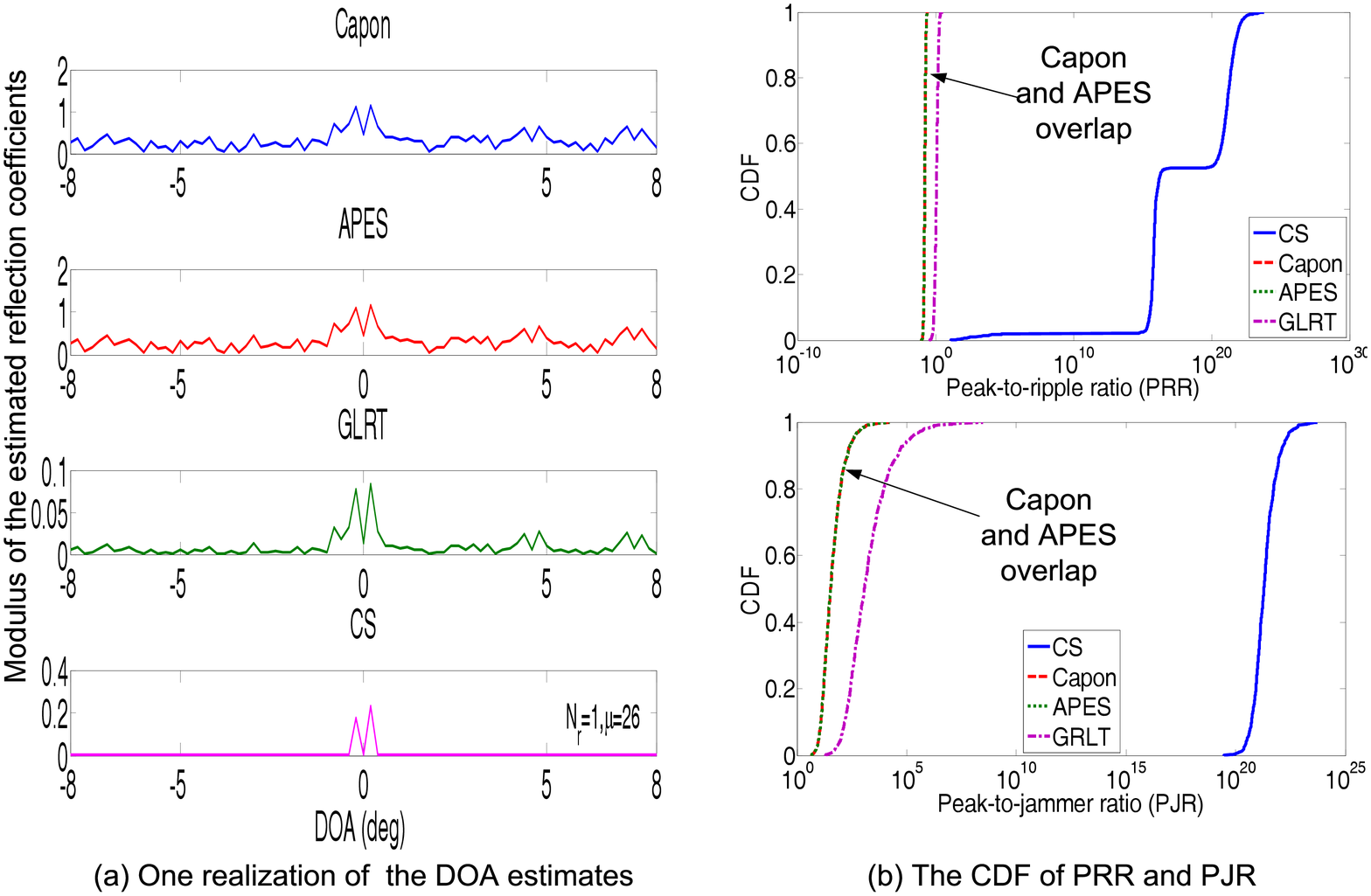}
 \caption{One realization of the DOA estimates (left column) and  CDF of PPR and PJR (right column).
 $N_r=1$,\ $M_t=M=30$,\ $\beta^2=400$,\ SNR$=0$ dB and $\mu=26$.
  }\label{PPR_com_five_betaj_20_Mt_30}
\end{figure}


\begin{figure}[htbp]
  \centering
    \includegraphics[height=3.8in,width=3.5in,clip=true]{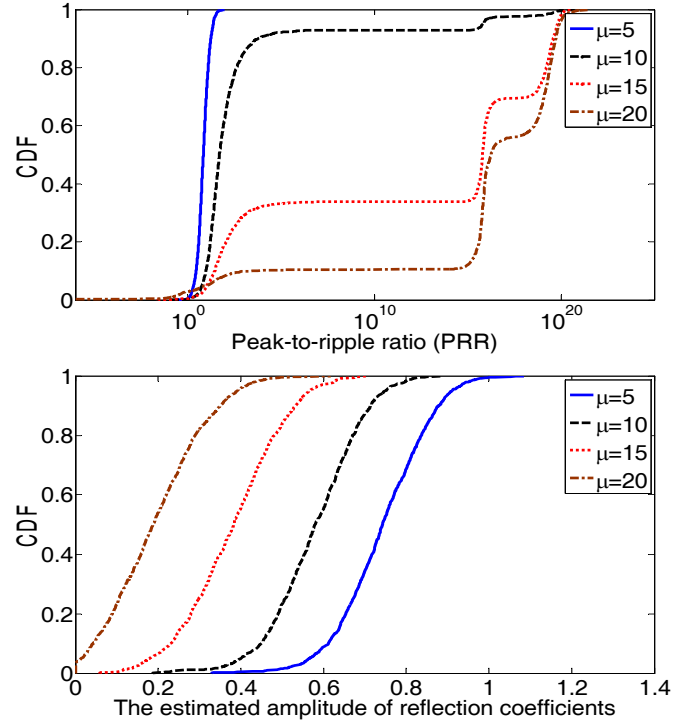}
 \caption{CDF of  PRR (top) and  amplitude estimate of RCS  (bottom).
  }\label{cdf}
\end{figure}

\begin{figure}[htbp]
  \centering
    \includegraphics[height=4in,width=5in,clip=true]{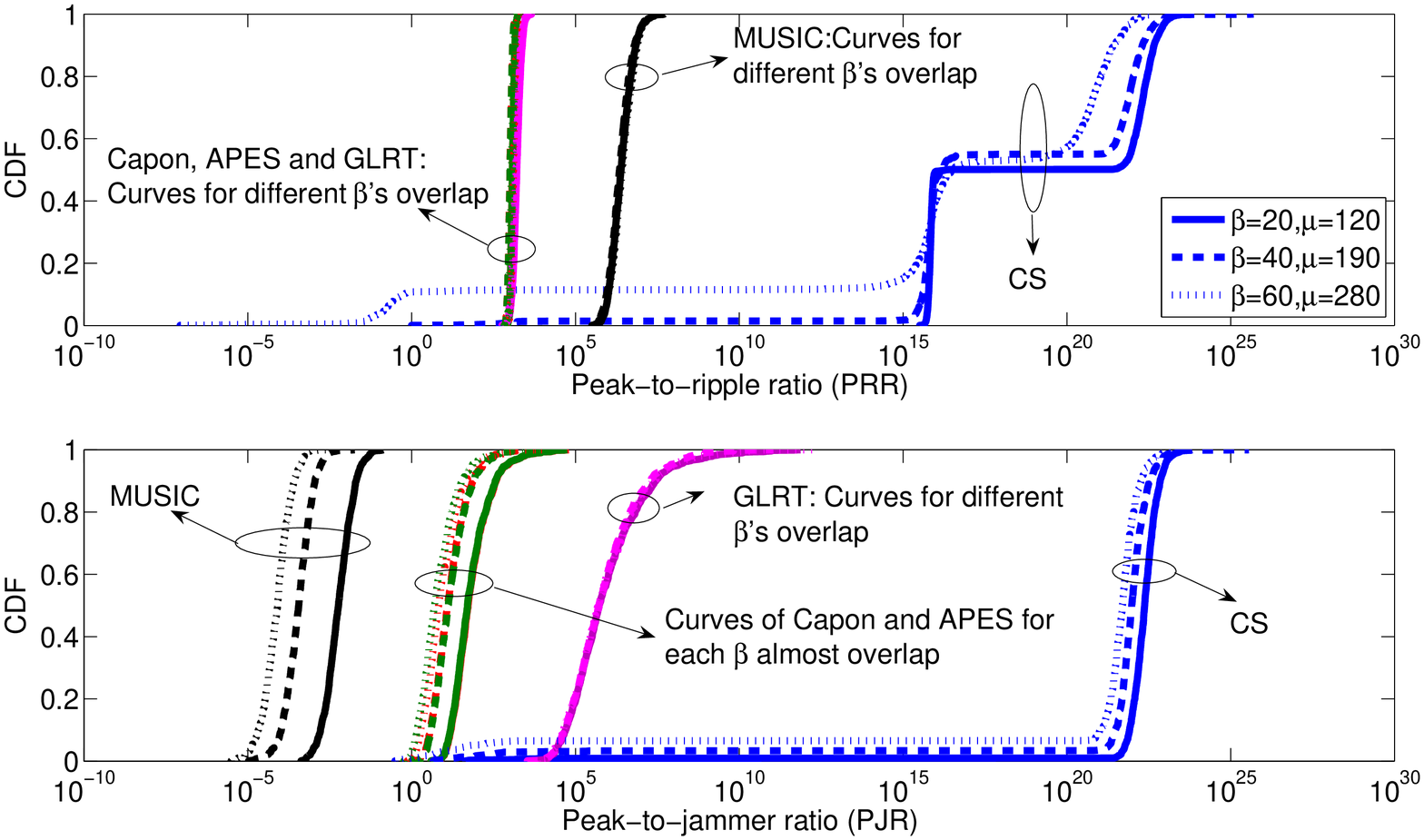}
 \caption{ CDF of  PRR and SJR for $\beta=20,40, 60$ with $N_r=10, M_t=M=30$ and SNR$=0$ dB.  The corresponding thresholds are $\mu=120,190$ and $280$. }\label{com_jammer}
\end{figure}

%

\begin{figure}[htbp]
  \centering
    \includegraphics[height=4in,width=5in,clip=true]{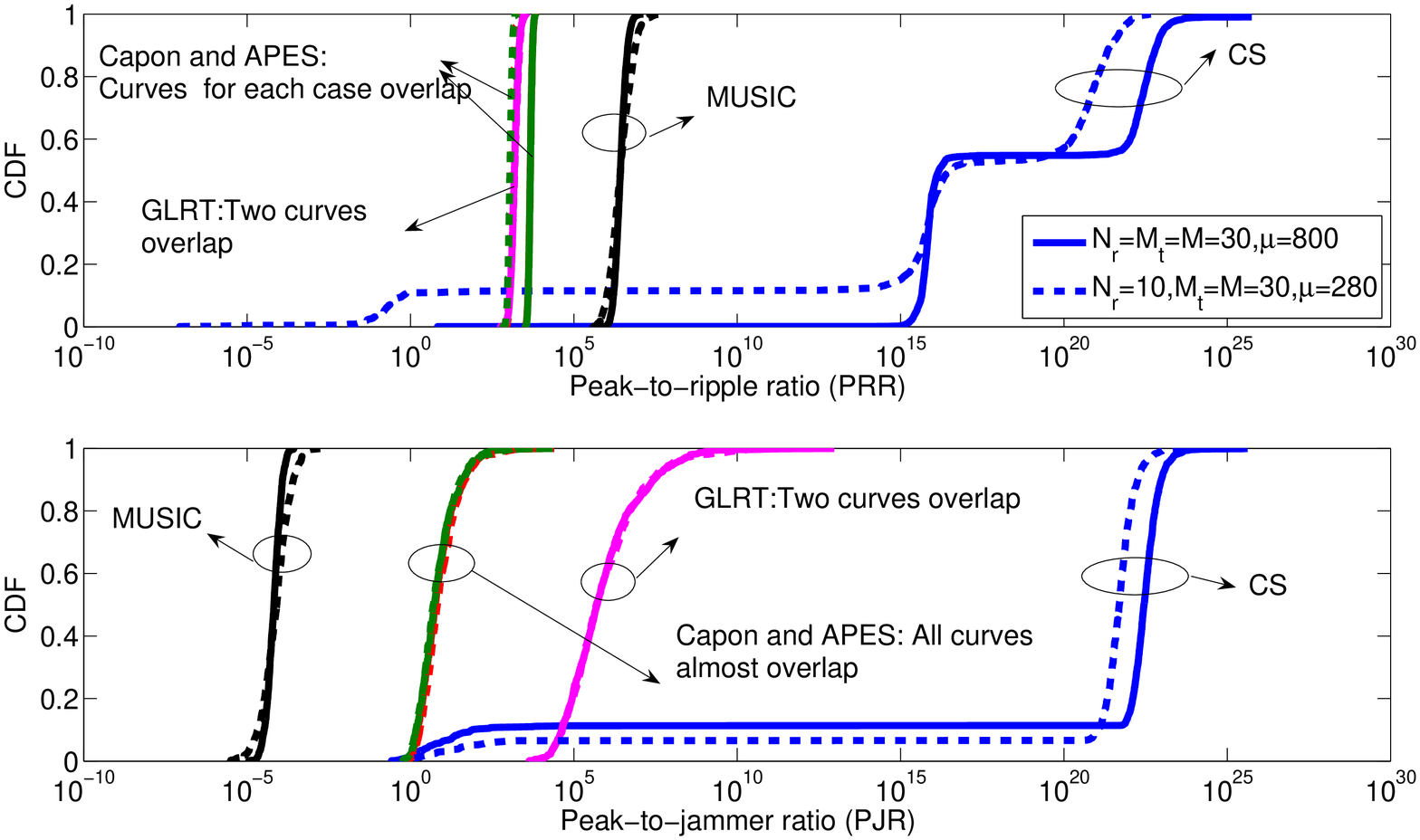}
 \caption{ CDF of  PRR and SJR for $\beta^2=3600$  and SNR=$0$ dB. Two cases are shown, $(N_r=10,M_t=30, M=30)$ and $(N_r=30,M_t=30, M=30)$.  The corresponding thresholds are $\mu=280$ and $800$. }\label{com_five_betaj_60_all_combination}
\end{figure}

\begin{figure}[htbp]
\centering
    \includegraphics[height=4in,width=5in,clip=true]{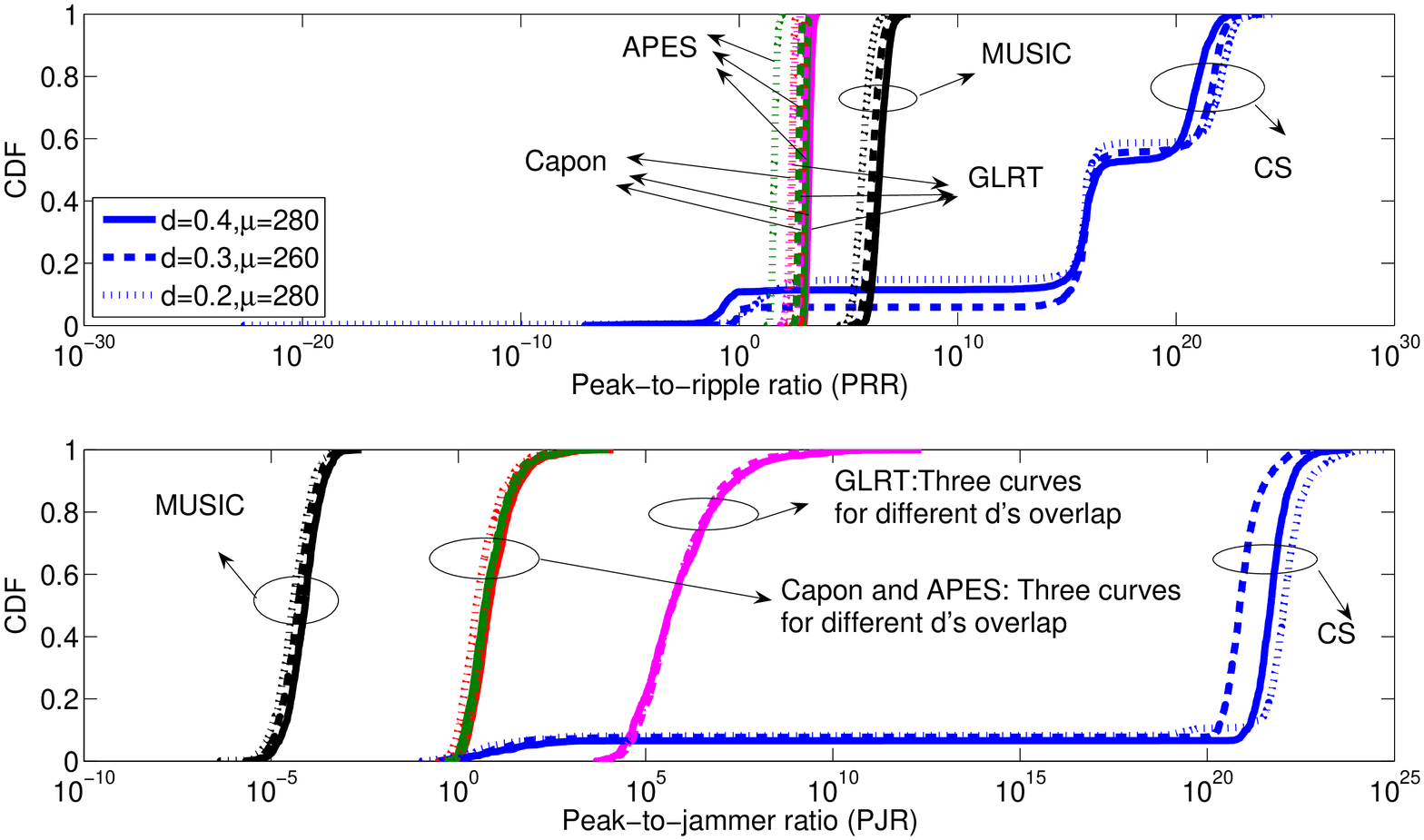}
 \caption{DOA estimates of two  targets with spacing $d=0.4\textordmasculine,0.3\textordmasculine$ and $0.2\textordmasculine$.    $N_r=10, M_t=M=30$, SNR=$0$ dB and  $\beta^2=3600$.
 The corresponding thresholds are $\mu=280,260$ and $280$.
  }\label{com_spacing}
\end{figure}


\begin{figure}[htbp]
\centering\includegraphics[height=5in,width=6in,clip=true]
{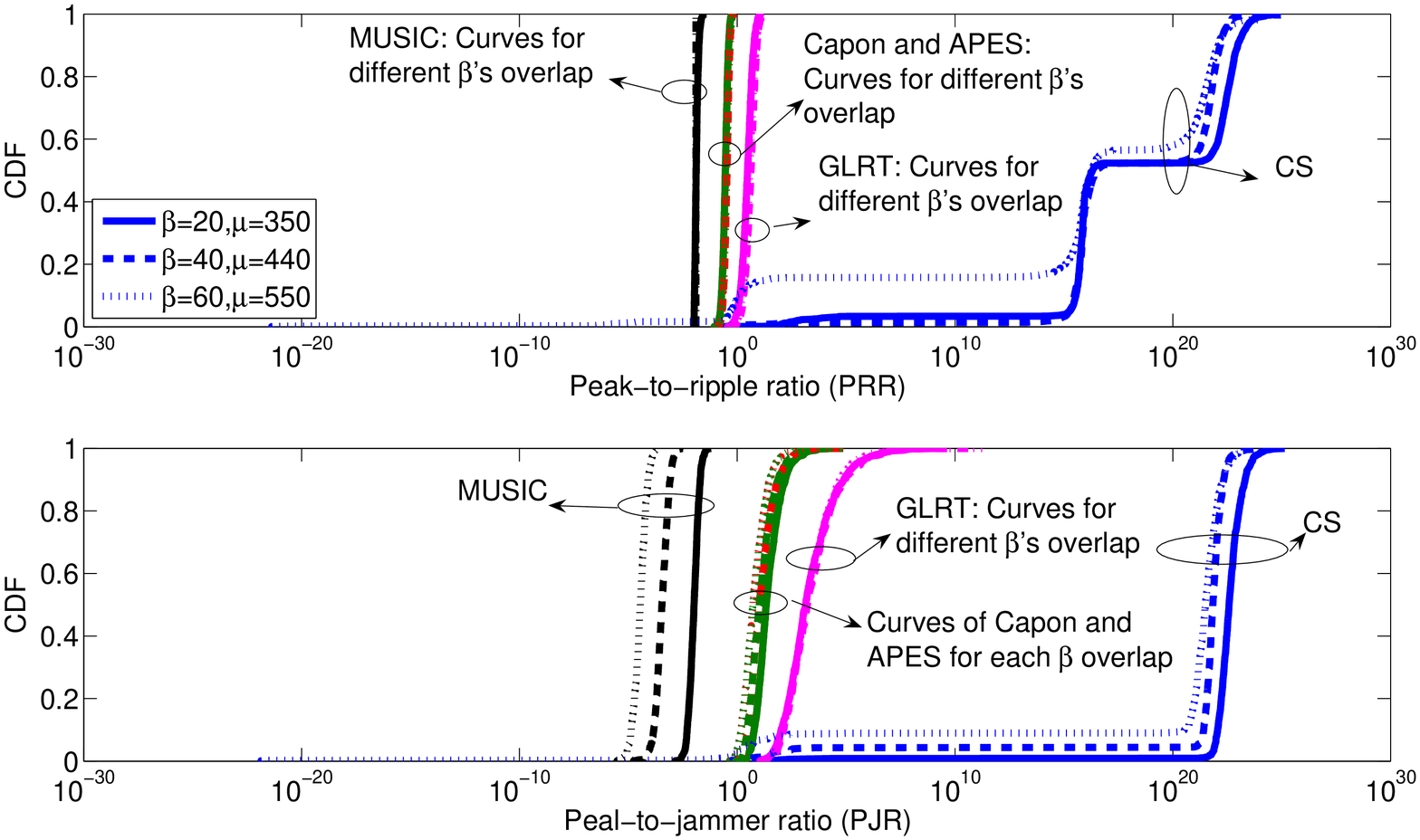}
 \caption{CDF of  PRR and PJR for $\beta=20,40$ and  $60$ with $N_r=20, M_t=M=30$ and SNR=$-40$ dB.  The corresponding thresholds are $\mu=350,440$ and $550$.
  }\label{com_jammer_snr_-40}
\end{figure}

\begin{figure}[htbp]
\centering\includegraphics[height=5in,width=6in,clip=true]{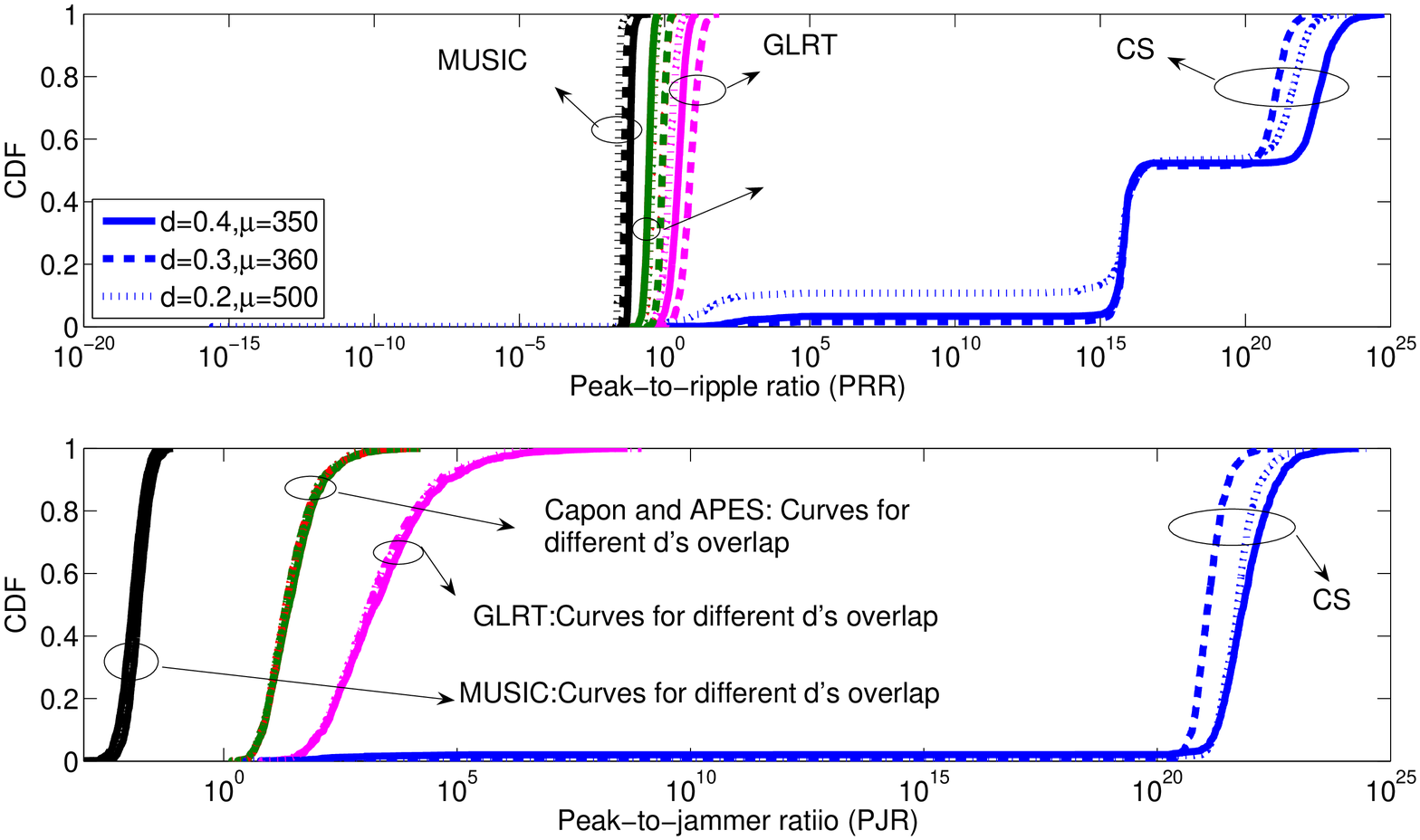}
 \caption{CDF of PRR and PJR for two targets with the spacing
 $d=0.4\textordmasculine,0.3\textordmasculine$ and $0.2\textordmasculine$.
 $N_r=20, M_t=M=30$,
 SNR=$-40$ dB and $\beta=20$. }\label{com_spacing_snr_-40}
\end{figure}

\begin{figure}[htbp]
\centering\includegraphics[height=4in,width=6in,clip=true]{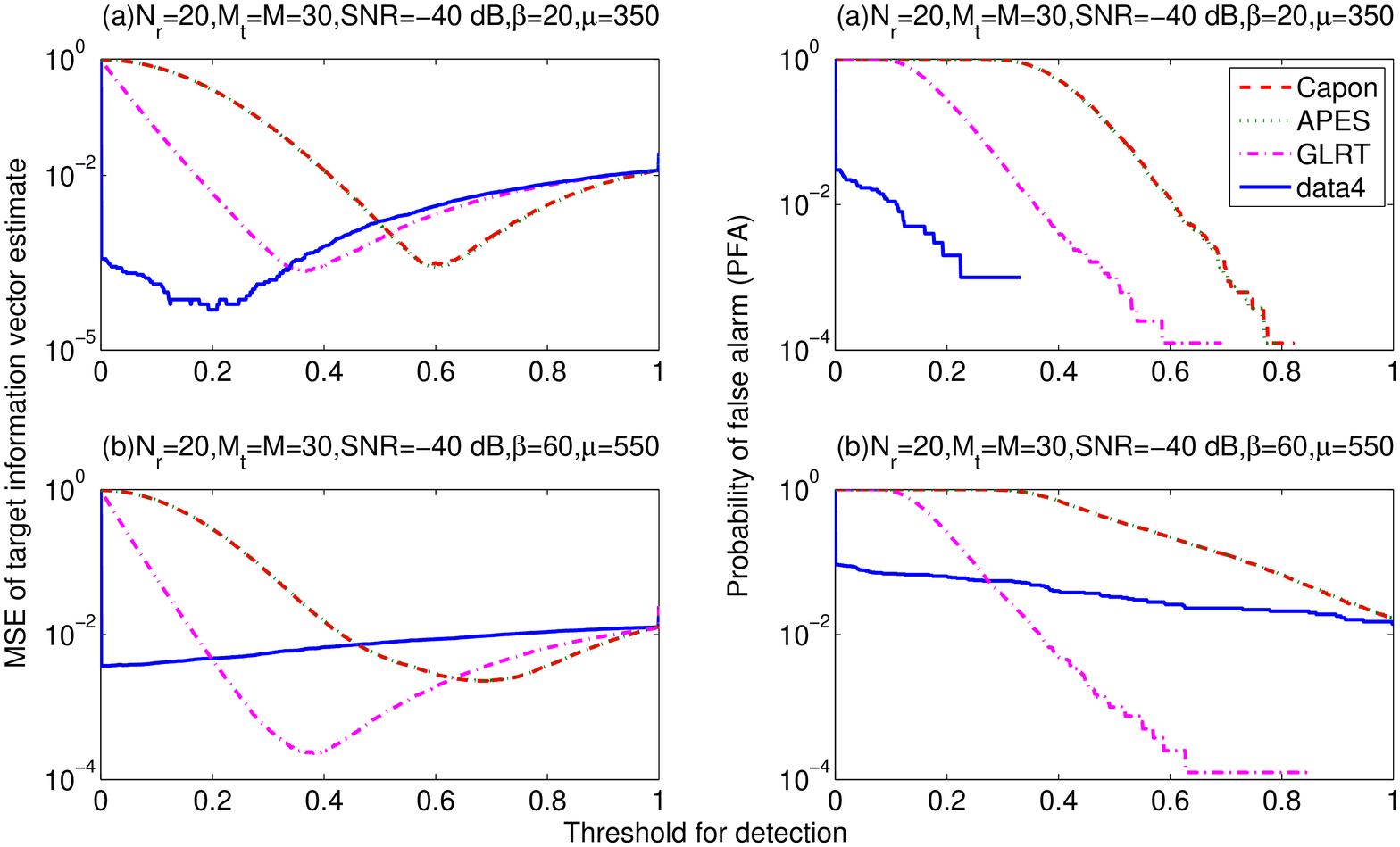}
 \caption{MSE of target information vector and probability of false alarm (PFA)   for two targets with spacing
 $d=0.4 \textordmasculine$ for $N_r=20,M_t=M=30$ and SNR$=-40$ dB.
 }\label{mse_pfa}
\end{figure}


\begin{figure}[htbp]
\centering\includegraphics[height=4in,width=6in,clip=true]{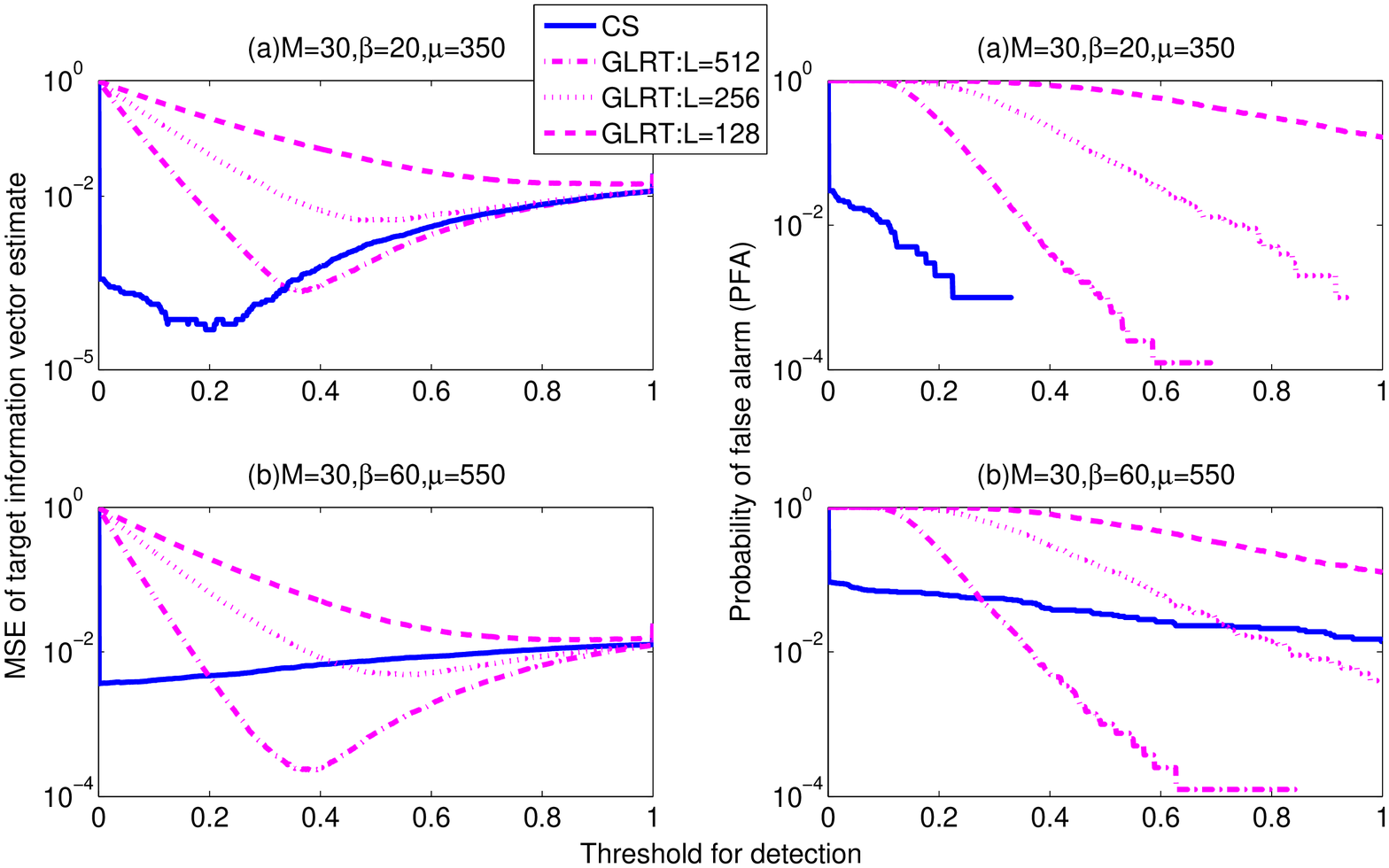}
 \caption{MSE of target information vector and probability of false alarm (PFA)   for two targets with spacing
 $d=0.4 \textordmasculine$ for $N_r=20,M_t=M=30$ and SNR$=-40$ dB.
 The number of transmit waveforms and receive samples per pulse for CS is 512 and 30, respectively.
 }\label{mse_pfa_new}
\end{figure}

%
%

\begin{figure}[htbp]
\centering\includegraphics[height=5in,width=6in,clip=true]
{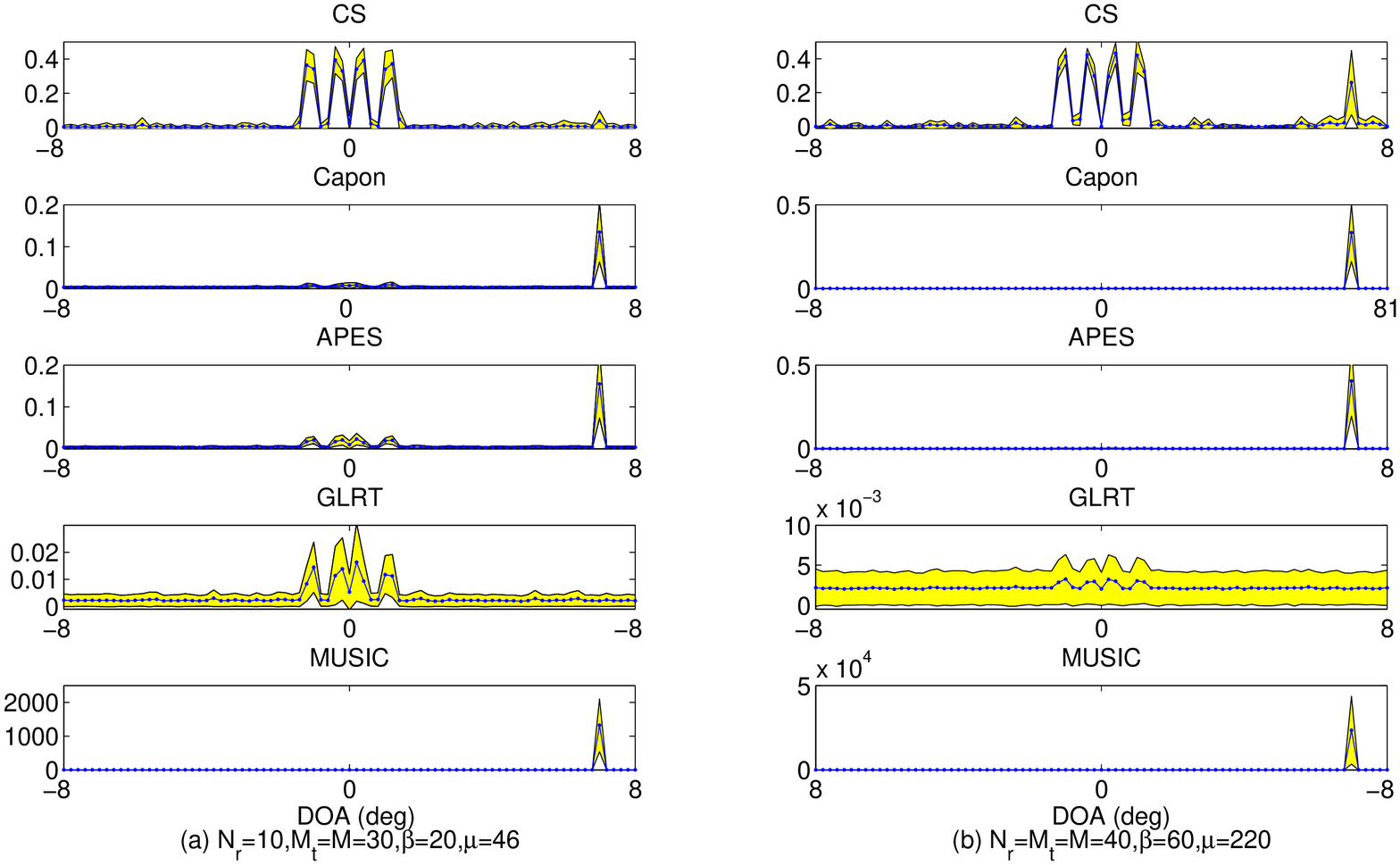}
 \caption{Modulus of DOA estimates for four targets that do not fall on  grid
 points. The dotted line is the mean of DOA estimates. The yellow
 region is the area bounded by the curves  mean $\pm$ std.
  }\label{com_angle_off_grid}
\end{figure}

\begin{figure}[htbp]
\centering
\includegraphics [height=3.5in,width=4.5in,clip=true] {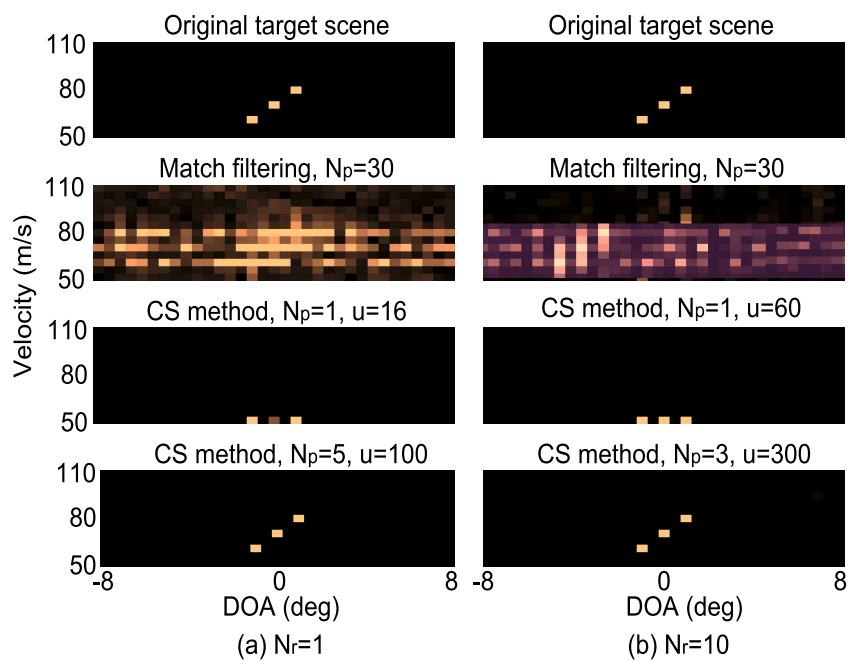}
\caption{Angle-Doppler estimates  for three targets on the grid
points. The three targets are located at \{-1\textordmasculine,
0\textordmasculine, 1\textordmasculine\}. $M_t=M=30$, SNR$=0$ dB and
$\beta^2=400$.
  }\label{angle_doppler_on_grid}
\end{figure}


\begin{figure}[htbp]
\centering
\includegraphics [height=3.5in,width=4.5in,clip=true] {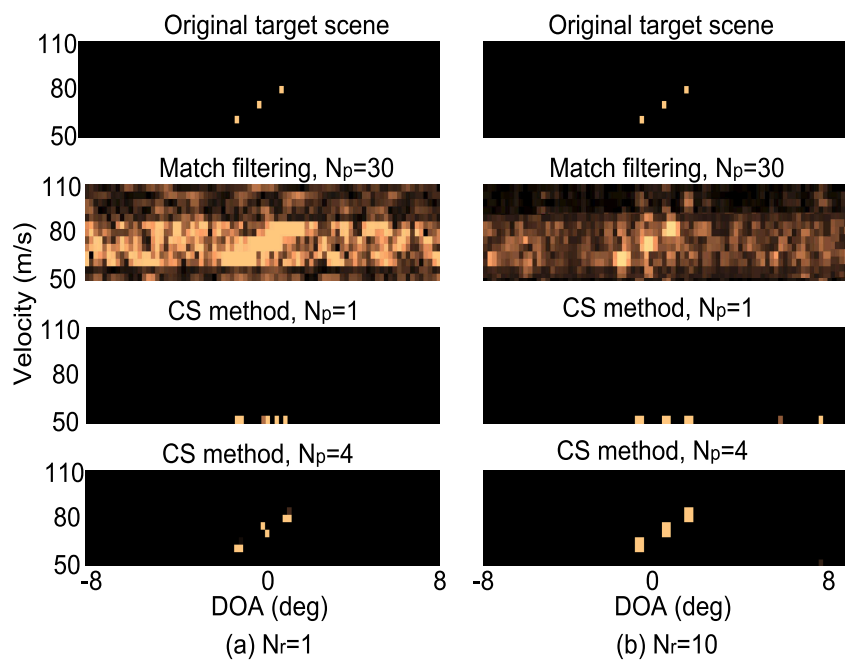}
\caption{Angle-Doppler estimates for three targets that do not fall
on the grid points. The three targets are located at
\{-1.1\textordmasculine, 0.1\textordmasculine,
1.1\textordmasculine\}. $M_t=M=30$, $\beta^2=400$ and SNR$=0$ dB.
  }\label{angle_doppler_off_grid}
\end{figure}


\begin{figure}[htbp]
\centering
\includegraphics [height=3.5in,width=4.5in,clip=true] {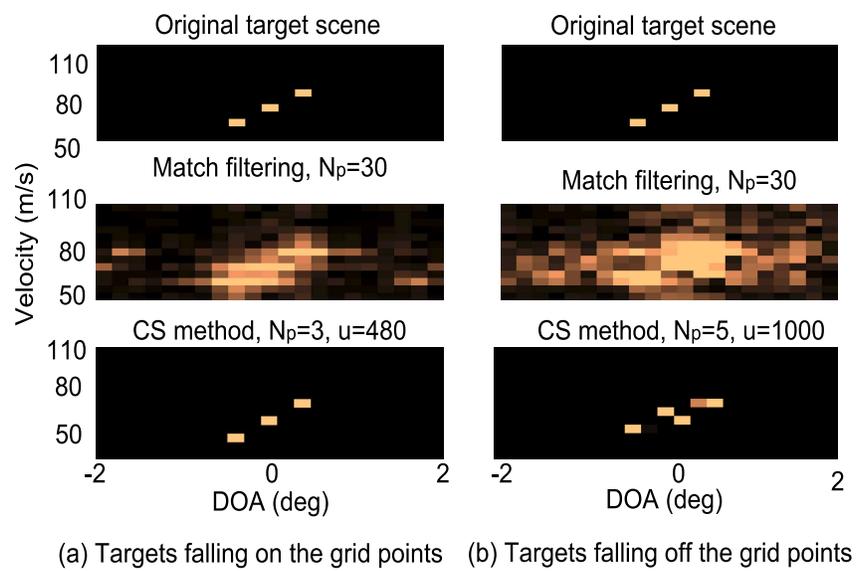}
\caption{Angle-Doppler estimates for three targets on and off grid
points. $N_r=10$, $M_t=M=30$, SNR$=0$ dB, $\beta^2=400$ and
$d=0.4\textordmasculine$.
  }\label{angle_doppler_on_grid_snr_0_d_0.4}
\end{figure}

\end{document}